\documentclass[11pt,a4paper]{article}
\usepackage{jcappub}
\usepackage{color,graphicx,amssymb,epsfig}

\def\Br{\boldsymbol B(\boldsymbol r)}
\def\Bir{\boldsymbol B_{i}(\boldsymbol r)}
\def\Bjr{\boldsymbol B_{j}(\boldsymbol r)}
\def\Bzr{\boldsymbol B_{0}(\boldsymbol r)}
\def\Bk{\boldsymbol B(\boldsymbol k)}
\def\B{{\boldsymbol B}}

\renewcommand{\vec}[1]{\boldsymbol{#1}}
 \definecolor{Black}{named}{Black}
 \definecolor{Blue}{named}{Blue}
 \definecolor{Red}{named}{Red}

\def\la{\mathrel{\mathpalette\fun <}}
\def\ga{\mathrel{\mathpalette\fun >}}
\def\fun#1#2{\lower3.6pt\vbox{\baselineskip0pt\lineskip.9pt
  \ialign{$\mathsurround=0pt#1\hfil##\hfil$\crcr#2\crcr\sim\crcr}}}

\title{Cosmic Ray Anisotropy as Signature for the Transition from Galactic to Extragalactic Cosmic Rays}

\author[a,b]{G.~Giacinti,}
\author[a]{M.~Kachelrie\ss,}
\author[c,d]{D.~V.~Semikoz,}
\author[b]{G.~Sigl}

\affiliation[a]{Institutt for fysikk, NTNU, Trondheim, Norway}
\affiliation[b]{II. Institut f\"ur Theoretische Physik, Universit\"at Hamburg, Germany}
\affiliation[c]{AstroParticle and Cosmology (APC), Paris, France}
\affiliation[d]{Institute for Nuclear Research of the Russian Academy of Sciences, Moscow, Russia}

\abstract{We constrain the energy at which the transition from Galactic to extragalactic cosmic rays occurs by computing the anisotropy at Earth of cosmic rays emitted by Galactic sources. Since the diffusion approximation starts to loose its validity for $E/Z \gtrsim 10^{16-17}$\,eV, we propagate individual cosmic rays using Galactic magnetic field models and taking into account both their regular and turbulent components. The turbulent field is generated on a nested grid which allows spatial resolution down to fractions of a parsec. Assuming sufficiently frequent Galactic CR sources, the dipole amplitude computed for a mostly light or intermediate primary composition exceeds the dipole bounds measured by the Auger collaboration around $E \approx 10^{18}$\,eV. Therefore, a transition at the ankle or above would require a heavy composition or a rather extreme Galactic magnetic field with strength $\gtrsim 10\,\mu$G. Moreover, the fast rising proton contribution suggested by KASCADE-Grande data between $10^{17}$\,eV and $10^{18}$\,eV should be of extragalactic origin.
In case heavy nuclei dominate the flux at $E \gtrsim 10^{18}$\,eV, the transition energy can be close to the ankle, if Galactic CRs are produced by sufficiently frequent transients as e.g.\ magnetars.
}


\begin{document}

\maketitle
%
%
%
\section{Introduction}
\label{Introduction}

The question at which energy the transition from Galactic to extragalactic cosmic rays (CRs) takes place is one of the major unresolved issues of cosmic ray physics. Two promising possibilities are to associate the transition with one of the two evident features of the cosmic ray spectrum: The second knee around $E\simeq 5\times10^{17}$\,eV or the ankle at $E\simeq3\times 10^{18}$\,eV. Since the chemical composition of galactic and extragalactic CRs should differ in general, both because of propagation effects and of the different nature of their sources, the transition may be detected experimentally studying the chemical composition of CRs as function of energy.

In the case of a transition around the second knee, Galactic CR sources such as e.g.\ supernova remnants would accelerate CRs up to the rigidity-dependent knee, which is close to $10^{17}$\,eV for iron. If the extragalactic CR flux dominating at higher energies would consist mainly of protons, the ankle could be explained as a dip in the extragalactic CR spectrum due to the pair-production losses of protons on cosmic microwave background (CMB) photons $p + \gamma_{\rm CMB} \rightarrow p + e^+ + e^-$~\cite{dip}. Below $\sim10^{17-18}$\,eV, the extragalactic CR flux may be suppressed because of CR propagation in extragalactic magnetic fields~\cite{Lemoine:2004uw,Kotera:2007ca}. On the other hand, the scenario of Ref.~\cite{Allard:2005cx} would favour a transition at the ankle. The composition of the CR flux at high energies is the subject of current debate due to the facts that hadronic physics must be extrapolated from lower energies and that the complex experimental analyses for different experiments are not yet completely reconciled. The scenario of Ref.~\cite{dip} is supported by the composition measurements of HiRes~\cite{HiRes} and the first results of the Telescope Array~\cite{TA}, which are consistent with a light composition around the ankle and above. On the other hand, recent results from the Pierre Auger Observatory~\cite{Abraham:2010yv,auger:2011pe} indicate a composition becoming heavier with increasing energy above the ankle, and the Yakutsk EAS array muon data suggests a non negligible fraction of heavy nuclei above $\simeq10^{19}$\,eV~\cite{Yakutsk}. Moreover, the measurements of the KASCADE-Grande~\cite{KASCADE-Grande} collaboration are consistent with a dominantly heavy composition up to $10^{18}$\,eV. However, the KASCADE-Grande data indicate a fast rising proton contribution above $10^{17}$\,eV.

Thus at present the experimental data on the CR composition do not allow us yet to determine the transition energy between Galactic and extragalactic CRs. In this paper we suggest to use instead experimental limits on the anisotropy of the arrival directions of UHECRs to constrain the maximal contribution of Galactic CRs at $E\gtrsim 10^{18}$\,eV. At energies below $10^{17}$\,eV, the diffusive propagation of Galactic cosmic rays and their resulting anisotropy at Earth was studied in details in Refs.~\cite{Blasi:2011fi,Blasi:2011fm}.

Since the propagation of CRs in the Galactic magnetic field (GMF) is not longer diffusive at $E\gtrsim 10^{17}$\,eV, we directly propagate UHECRs in the GMF using the numerical code developed in Refs.~\cite{Giacinti:2010dk,Giacinti:2011uj}. We present also a way to generate the turbulent field on a nested grid without limitation on its spatial resolution. This method allows us to include magnetic field fluctuations spanning the required large dynamical range of scales, from negligible compared to the CR Larmor radii up to 300\,pc. As main result of this work we show that the existing limits on CR anisotropies strongly restrict the contribution of the CNO element group to the Galactic CR component above $E\gtrsim1$\,EeV, while the contribution of iron is restricted above $E\gtrsim3$\,EeV.

Details of the method to generate turbulent magnetic fields are discussed in the Section~\ref{Method_TF}. In Section~\ref{Method_Anisotropy}, we review the GMF models used and discuss how the CR anisotropy is calculated.
Results of numerical simulations are presented in the Sections~\ref{Anisotropy} and \ref{Spectrum} for anisotropies and the spectrum of UHECR.

%
%
%

\section{Modeling Turbulent Magnetic Fields}
\label{Method_TF}

We adopt in this section a convenient way to generate turbulent magnetic fields on nested grids which allows to include a large dynamic range of spatial scales contributing to the turbulence.

A turbulent magnetic field $\B$ satisfies $\left\langle\Br\right\rangle=\textbf{0}$ and
$\left\langle\Br^2\right\rangle \equiv B_{\rm rms}^2>0$. Let us denote $k$ the modulus
of wave vectors and $\alpha$ the spectral index of the field: $\alpha = 5/3,\,3/2$ and 1 respectively for Kolmogorov, Kraichnan and Bohm spectra. The power spectrum of the field satisfies $\mathcal{P}(k)\propto k^{-\alpha}$, and the amplitudes of its Fourier modes are $|\Bk|^2\propto k^{-\alpha -2}$. The spectral index of the turbulent Galactic magnetic field is poorly constrained. While $\alpha \simeq 1$ appears hardly plausible, both Kolmogorov and Kraichnan spectra could be allowed by the data. To study the dependence of our results on the spectral index, we present below computations for $\alpha = 5/3$ and $3/2$, as examples. Wave vector moduli satisfy $2\pi/L_{\max}\le k=|\textbf{k}|\le2\pi/L_{\min}$, where $L_{\min}$ and $L_{\max}$ are respectively the minimal and the maximal variation scales in the turbulent field. In practice, $L_{\min}$ corresponds to the damping scale of the field, which could be as low as an astronomical unit. We choose here $L_{\min} = 1$\,AU. For $\alpha = 5/3$ and $3/2$, the value of $L_{\min}$ does not noticeably affect the results, because the larger $\alpha$ is, the more the energy is concentrated in the modes with large spatial variations. We take $L_{\max} = 100 - 300$\,pc. The correlation length $L_{\rm c}$ of the field, defined as in~\cite{Harari:2002dy}, is equal to
\begin{equation}
L_{\rm c} = \frac{L_{\max}}{2} \: \frac{\alpha - 1}{\alpha} \: \frac{1-(L_{\min}/L_{\max})^{\alpha}}{1-(L_{\min}/L_{\max})^{\alpha-1}}~.
\label{CorrelationLength}
\end{equation}

As discussed in Refs.~\cite{DeMarco:2007eh,Giacinti:2011uj}, there are two main numerical methods to generate turbulent magnetic fields. First, they can be generated as a superposition of plane waves as in Ref.~\cite{JG99} and computed in any point of the space. Second, values of the field can be pre-computed with the Fast Fourier Transform on a three dimensional cubic grid, which is periodically repeated in space. The value of the field can be extrapolated to any position from these values. Computing the individual trajectories of millions of cosmic rays with rigidities as low as $E/Z \sim 3 \times 10^{16}$\,eV is achievable within reasonable computing times only with the second method. The number of vertices on such cubic grids is $\mathcal{N}^3$. $\mathcal{N} \sim 256 - 512$ is typically the limit above which the grid cannot be loaded in a 2~gigabyte RAM memory. The ratio $L_{\max}/L_{\min}$ is limited by $\mathcal{N}/2$, when $L_{\max}$ equals the size of the cubic box. Moreover, we take $L_{\max}/L_{\min}$ to be smaller than $\mathcal{N}/2$, by at least a factor of a few. This ensures that the modes with the largest spatial variations $\simeq L_{\max}$ have a few oscillations within the box size. Otherwise, the generated turbulent field can be highly anisotropic. Cosmic rays which diffuse in turbulent magnetic fields are mostly sensitive to modes with wave numbers $k$ close to $\sim 2\pi/r_{\rm L}$, where $r_{\rm L}$ is their Larmor radius. For $E/Z = 10^{18}/26$\,eV and a field of strength $6\,\mu$G, it is equal to $r_{\rm L}\simeq 7$\,pc. In the numerical simulations, one can disregard modes with $2\pi /k \ll r_{\rm L}$ because they have a negligible influence on the particle trajectories. On the contrary, modes with $2\pi /k \in [r_{\rm L},L_{\max}]$ which isotropize cosmic rays in a non trivial way have to be taken into account. Therefore, instead of using $2\pi /k \in [L_{\min},L_{\max}]=[1\,{\rm AU},100-300\,{\rm pc}]$, we truncate the minimal scale of spatial variations for the generated field and restrict ourselves to $[L_{\min}',L_{\max}]$ with $L_{\min}'$ sufficiently small compared to $r_{\rm L}$. However, $L_{\max}/L_{\min}'$ is still too large to fit in one magnetic field grid of reasonable size. To solve this issue, we use the method of nested grids explained in the following.

Let us assume that $\Br$ is the sum of $N+1$ components: $\Br = \sum_{i=0}^{N} \Bir$ (for $j \neq i$, $\left\langle \Bir \cdot \Bjr \right\rangle = 0$). In practice, $N=2$ is sufficient for this work. $\Bzr$ contains all Fourier modes with $2\pi/k \in [L_{\min},L_{\min}']$, and the fields $\Bir$ ($1 \leq i \leq N$) respectively contain the modes with $2\pi/k \in [L_{i},L_{i+1}]$, where $L_1 = L_{\min}'$ and $L_{N+1} = L_{\max}$. The ratios $L_{i}/L_{i+1}$ are all chosen to be smaller than $\mathcal{N}/2$ by a factor of a few.

The root mean square (rms) strength of the total field, $B_{\rm rms}$, satisfies~\cite{Tinyakov:2004pw}
\begin{equation}
B_{\rm rms}^2 \propto \int_{2\pi/L_{\max}}^{2\pi/L_{\min}} dk \: \mathcal{P}(k)~.
\label{Brms}
\end{equation}
Therefore, for $\alpha \neq 1$, $B_{\rm rms}^2 \propto (L_{\max}^{\alpha -1} - L_{\min}^{\alpha -1})$. For $0 \leq i \leq N$, the energy density present in $\Bir$ is proportional to
\begin{equation}
B_{{\rm rms}, i}^2 \propto B_{\rm rms}^2 \: \frac{L_{i+1}^{\alpha -1} - L_{i}^{\alpha -1}}{L_{\max}^{\alpha -1} - L_{\min}^{\alpha -1}}~,
\label{BrmsPrime}
\end{equation}
which yields the rms amplitude of $\Bir$, $B_{{\rm rms}, i}$. The turbulent field $\boldsymbol B_{\rm turb}$ generated for the computations is equal to the sum of the $N$ components $\Bir$ with $i=1,...,N$, $\boldsymbol B_{\rm turb}=\sum_{i=1}^{N} \Bir$. $\boldsymbol B_{\rm turb}$ is equal to the total turbulent field $\Br$ after subtracting the modes with spatial variation scales smaller than $L_{\min}' \ll r_{\rm L}$. Each $\Bir$ is generated on a cubic grid of lateral size $\mathcal{N}L_{i}/2$. Each grid is periodically repeated in physical space. The $\Bir$ with large $i$ contain the modes with large spatial variation scales and the $\Bir$ with small $i$, the modes with small variation scales. In any space point, the magnetic field from the large (respectively small) resolution grid is evaluated as the 8-point linear interpolation of the values on vertices of the large (respectively small) scale resolution grid.

We have verified that we recover with this code the results found by the earlier studies of Refs.~\cite{DeMarco:2007eh,Casse:2001be}. As an example, we present in the appendix our computations of the CR diffusion coefficient for pure magnetic turbulence, as well as the parallel and perpendicular diffusion coefficients for turbulence superimposed to a regular field. Our results are found to be in very good agreement with those of these previous studies.

To summarize, we take in the following $L_{\min}=1$\,AU for the normalisation of the turbulent field strength, so that the rms for the total field with modes satisfying $2\pi /k \in [L_{\min}=1\,{\rm AU},L_{\max}]$ would be $B_{\rm rms}$. In all computations for CRs with rigidities $E/Z<1$\,EeV, we take $N=2$ and set $L'_{\min}$, the actual minimal scale of fluctuations in the generated field, to 1\,pc. The intermediate scale $L_2$ between the $N=2$ grids is 20\,pc. In practice, for large rigidities $E/Z \geq 1$\,EeV, $r_{\rm L} \geq 180$\,pc and we can drop the smaller scale grid and only use the larger one : $N=1$ and $L'_{\min}=20$\,pc. We use the standard Runge Kutta method with adaptative step size of Ref.~\cite{NR}. Removing the smaller scale grid increases the step size of the integrator and allows us to reduce computing time for particles with $E/Z\geq1$\,EeV.

%
%
%

\section{Galactic Magnetic Field models and Method to compute the Anisotropy}
\label{Method_Anisotropy}

The Galactic magnetic field (GMF) can be regarded as the sum of a regular component (large scale variations) and a turbulent component (small scale variations).

We described in the previous Section a method to generate numerically the turbulent component. The spatial profile of the rms strength of the turbulent field, $B_{\rm rms}(r,z)$, is poorly constrained. Therefore, we use two different types of profiles as examples. First, we take a model with an exponentially decaying field strength in the Galactic halo~\cite{Giacinti:2009fy}. We will refer to it as the ``Profile 1'':
\begin{equation} \label{profile1}
B_{\rm rms}(r,z) = B(r) \exp\left(-\frac{|z|}{z_{0}}\right)~,
\end{equation}
where $r$ is the Galactocentric radius and $z$ the distance to the Galactic plane. The parameter $z_{0}$ denotes
the scale height of the random field into the $z-$direction. We will take $z_{0}=(2-8)$\,kpc in this work. The radial profile $B(r)$ is equal to
\begin{equation}
B(r) = \left\{ \begin{array}{ll}
                B_{0} \: \exp{\left( \frac{5.5}{8.5} \right) }  & \mbox{, if } r \leq 3\,\mbox{kpc (bulge)}\\
		B_{0} \: \exp{\left( \frac{ - \left( r - 8.5\,{\rm kpc}\right)}{8.5\,{\rm kpc}} \right) } & \mbox{, if } r > 3\,\mbox{kpc}
               \end{array} \right.
\end{equation}
where $B_{0}$ is defined as the value of $B_{\rm rms}$ close to the Sun.

Second, we also consider a constant rms strength within a box of size $r \leq 20$\,kpc and $|z| \leq z_{0}$ (``Profile 2''):
\begin{equation}  \label{profile2}
B_{\rm rms}(r,z) = \left\{ \begin{array}{ll}
                B_{0} & \mbox{, if } r \leq 20\,\mbox{kpc and }|z| \leq z_{0}\\
		0 & \mbox{, if } r > 20\,\mbox{kpc or }|z| > z_{0}
               \end{array} \right.
\end{equation}
Although this profile is very likely less realistic than the previous one, we test it because it corresponds to the profile used in the usual ``leaky-box approximation''.

The global geometry of the regular GMF is still poorly known. The Faraday rotation measures (RM) for extragalactic sources suggest that it is made of at least two different components, in the disk and in the halo, with different geometries~\cite{Jansson:2009ip,Pshirkov:2011um}. The field in the disk is believed to be symmetric with respect to the Galactic plane, while the field in the halo is believed to be antisymmetric~\cite{Jansson:2009ip,Pshirkov:2011um}. The RM at high latitudes show the existence of a toroidal field in the halo, on each side of the Galactic plane. This field is counter clockwise in the Northern halo and clockwise in the Southern halo, as seen from the Galactic North pole. Several analytical models have been proposed to describe the regular GMF. As shown in Refs.~\cite{Jansson:2009ip,Waelkens:2008gp}, presently no theoretical GMF model can fit all experimental data. However, Ref.~\cite{Pshirkov:2011um} presents the two first models that fit reasonably well all extragalactic RM data in most regions of the sky, which represents a significant improvement of our knowledge of the GMF. Since it is impossible to distinguish between bisymmetric (BSS) and antisymmetric (ASS) geometries for the disk field, the authors of Ref.~\cite{Pshirkov:2011um} propose two different benchmark models for the regular GMF. Below, we will refer to them as the ``PTKN-BSS'' and ``PTKN-ASS'' models. They contain disk and toroidal contributions. Let us use Galactocentric cylindrical coordinates $(r,\theta,z)$, where $r=\left(x^{2}+y^{2}\right)^{1/2}$, and Cartesian coordinates $x$, $y$ and $z$. The Earth is assumed to be at $(x=0,\,y=r_{\odot}=8.5\,{\rm kpc},\,z=0)$, where $\theta$ is set to zero at the position of the Earth, and increases clockwise, as seen from the Galactic North pole. The components in cylindrical coordinates of the disk field strength, $B_{\rm r}$ and $B_{\theta}$, are defined as
\begin{equation}
\begin{array}{ll}
B_{\rm r} = B\left(r,\theta,z\right)\sin p,\\
B_{\theta} = B\left(r,\theta,z\right)\cos p\,,
\end{array}
\end{equation}
with $p=-5^{\circ}$ and $p=-6^{\circ}$ respectively for the ASS and BSS models. For the ASS disk field,
\begin{equation}
 B\left(r,\theta,z\right) = b\left(r\right) \left| \cos\left[\theta-\frac{1}{\tan p} \ln\left( \frac{r}{r_{\odot}}\right) + \phi \right] \right| \, \cdot\, \exp\left( -\frac{|z|}{z_{0}} \right)\,,
\end{equation}
while, for the BSS disk field,
\begin{equation}
 B\left(r,\theta,z\right) = b\left(r\right) \cos\left[\theta-\frac{1}{\tan p} \ln\left( \frac{r}{r_{\odot}}\right) + \phi \right] \, \cdot\, \exp\left( -\frac{|z|}{z_{0}} \right)\,,
\end{equation}
where $\phi=1/\tan p \cdot \ln ( 1 + d/r_{\odot} ) - \pi /2$ with $z_{0}=1.0$\,kpc, $d=-0.6$\,kpc and
\begin{displaymath}
b(r) = \left\{ \begin{array}{ll}
                2.0\,\mu{\rm G} \, \frac{r_{\odot}}{5.0\,{\rm kpc}\,\cos \phi} & \mbox{for }r\leq 5.0\,{\rm kpc}\\
		2.0\,\mu{\rm G} \, \frac{r_{\odot}}{r\,\cos \phi} & \mbox{for }r > 5.0\,{\rm  kpc}
               \end{array} \right.\,.
\end{displaymath}

The halo field components $B_{\rm Tx}$ and $B_{\rm Ty}$ are defined as
\begin{equation}
\begin{array}{ll}
 B_{\rm Tx} = -B_{\rm T}~\mbox{sgn}\left(z\right)\cos\theta,\\
 B_{\rm Ty} = B_{\rm T}~\mbox{sgn}\left(z\right)\sin\theta\,,
\end{array}
\end{equation}
where
\begin{equation}
B_{\rm T} = B_{\rm T0} \cdot \frac{\frac{r}{r_{\rm T0}} \exp\left( \frac{r_{\rm T0} - r}{r_{\rm T0}}\right) }{1+\left( \frac{|z|-h_{\rm T}}{w_{\rm T}}\right)^{2}}\,,
\end{equation}
with $B_{\rm T0}$, $r_{\rm T0}$, $h_{\rm T}$ chosen as in Table~\ref{ParameterValues}. For $|z| \leq h_{\rm T}$, $w_{\rm T}=0.25$\,kpc, and for $|z|>h_{\rm T}$, $w_{\rm T}=0.4$\,kpc. The strength of the halo field decays towards the Galactic center, for $r < r_{\rm T0}$.

For most of the following computations, we use the PTKN-BSS model as an example. We test the dependence of our results on the regular GMF by also using the PTKN-ASS model, the ``ASS+RING'' model of Ref.~\cite{Sun:2007mx} (which we will refer to as ``Sun08'' in the following), and the Prouza and Smida (PS) model~\cite{PS,Kachelriess:2005qm} with the parameters given in Ref.~\cite{Giacinti:2010dk}.

\begin{table}
\begin{center}
\begin{tabular}{|c|c|c||c|c|}
\hline
		& ASS, $z>0$	& ASS, $z<0$	& BSS, $z>0$	& BSS, $z<0$ \\
\hline
$B_{\rm T0}$	& 4\,$\mu$G	& 2\,$\mu$G	& 4\,$\mu$G	& 4\,$\mu$G \\
\hline
$r_{\rm T0}$	& 6\,kpc	& 6\,kpc	& 6\,kpc	& 5\,kpc \\
\hline
$h_{\rm T}$	& 1.3\,kpc	& 1.3\,kpc	& 1.5\,kpc	& 1.5\,kpc \\
\hline
\end{tabular}
\end{center}
\caption{Values for the Northern ($z>0$) and Southern ($z<0$) halo parameters $B_{\rm T0}$, $r_{\rm T0}$ and $h_{\rm T}$, in the ASS and BSS versions of the PTKN model.}
\label{ParameterValues}
\end{table}

The Galactic center~\cite{581441,astro-ph/0411471,astro-ph/0504323}, some types of supernovae~\cite{Ptuskin:2010zn}, magnetars~\cite{Venkatesan:1996jw,Blasi:2000xm,Arons:2002yj,astro-ph/0405310} or GRBs~\cite{Waxman:1995vg,Vietri:1995hs,Waxman:2000vc,Gialis:2003fm,astro-ph/0504158,Murase:2006mm,Murase:2008mr} have been discussed as potential Galactic sources able to accelerate CRs to ultra-high energies. The spatial extension of the region containing Galactic CR sources is better constrained than the GMF parameters. Sources are expected to be distributed in the Galactic disk, within $\simeq \pm $(200--500)\,pc from the Galactic plane $z=0$~\cite{Strong:1998pw,Cordes:1997my}. The Galactocentric radius $r$ up to which the source region extends is less constrained. Since there should not be a significant number of sources with $r>20$\,kpc~\cite{Strong:1998pw}, we will take in the following, for most cases, $r=20$\,kpc as the limit of the source region.

At sufficiently low rigidities, the Larmor radius of cosmic rays is smaller than the coherence length of the turbulent GMF. Previous studies that predicted the amplitude of the cosmic ray anisotropy at Earth assumed CRs are diffusive. While the diffusion approximation is justified for rigidities smaller than $E/Z \sim 10^{17}$\,eV, it starts to fail in the rigidity range investigated in this work: $E/Z \geq(10^{18}/26)$\,eV. Between these rigidities, one typically expects a transition from the diffusive regime to the ballistic regime for CR propagation. The transition does not happen abruptly at a given rigidity, which leads to non-trivial modes of CR propagation. This can have a non-trivial impact on the anisotropy of Galactic CRs at Earth. Therefore, we propagate in this work individual cosmic rays in models of the GMF.

\begin{figure}
\begin{center}
\includegraphics[width=0.49\textwidth]{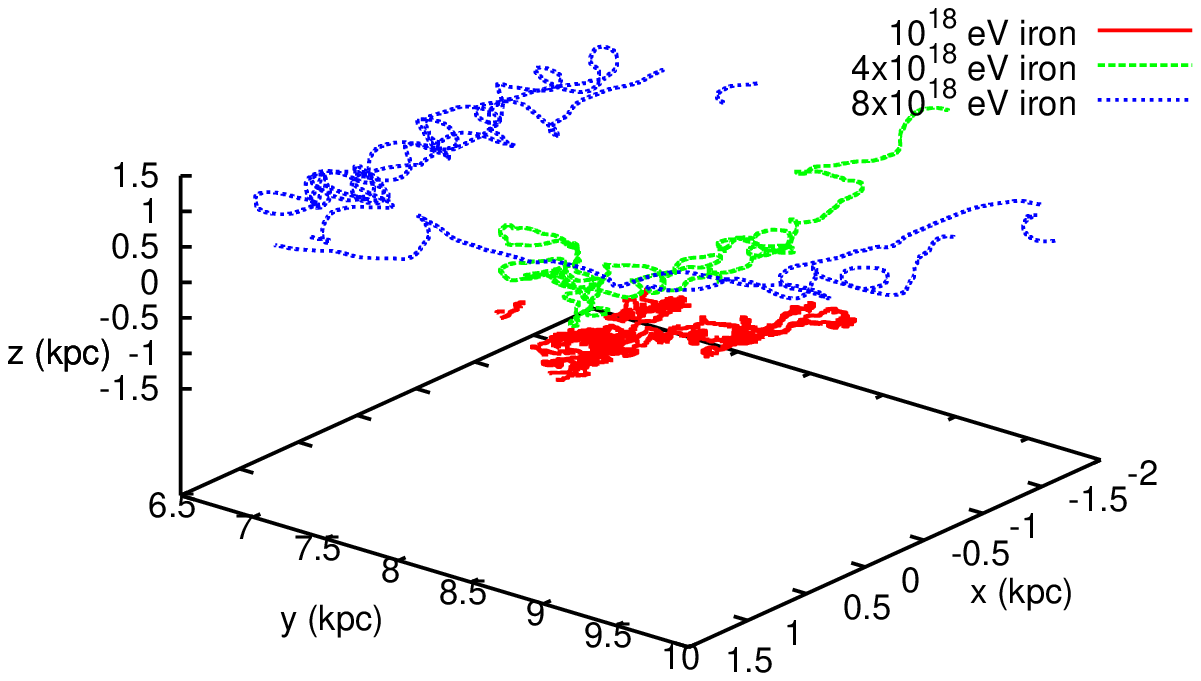}
\includegraphics[width=0.49\textwidth]{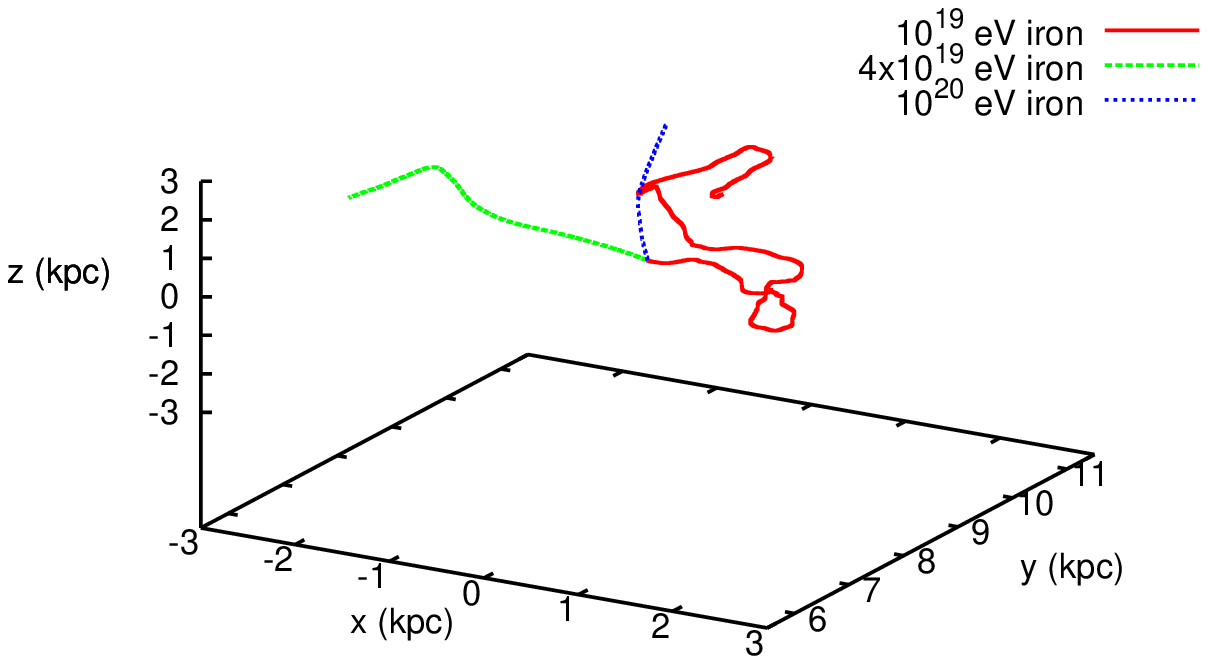}
\end{center}
\caption{Trajectories of iron anti-nuclei backtraced from the Earth. \textbf{Left panel:} Energies equal to $(1,4,8) \times 10^{18}$\,eV; \textbf{Right panel:} Energies equal to $(1,4,10) \times 10^{19}$\,eV. For details on the Galactic magnetic field model, see text.}
\label{Trajectories}
\end{figure}

Figure~\ref{Trajectories} shows trajectories of anti-iron nuclei with energies $10^{18}\,{\rm eV} \leq E \leq 10^{20}\,{\rm eV}$ backtraced in one GMF model from the Earth, located at ($x=0$,\,$y=8.5$\,kpc,\,$z=0$). It shows the variety of CR propagation types in the transition from ``purely'' diffusive (here at $10^{18}$\,eV) to ``purely''ballistic (here above $\ga (2-4) \times 10^{19}$\,eV). This energy range is shifted when the magnetic field parameters are changed. The regular GMF used for these plots is the PTKN-BSS model. The turbulent component has a Kolmogorov spectrum with $L_{\min}=1$\,AU, $L_{\max}=200$\,pc, the profile 1 with $z_0=2$\,kpc, and a strength set to $B_{\rm rms}=4\,\mu$G.

The left panel of Figure~\ref{Trajectories} display the trajectories of 1, 4 and 8 $\times 10^{18}$\,eV iron anti-nuclei. Values of spatial coordinates on the axes are given in kilo-parsecs. For these GMF parameters, the Larmor radius of the $10^{18}$\,eV nuclei is smaller than the correlation length $L_{\rm c}\simeq 40$\,pc of the turbulent component. The trajectory of this cosmic ray resembles a random walk, see the red line. The green ($4\times 10^{18}$\,eV) and blue ($8\times 10^{18}$\,eV) trajectories, respectively, correspond to diffusion in the regimes when $r_{\rm L} \simeq L_{\rm c}$ and $r_{\rm L} > L_{\rm c}$. The trajectory of the $8\times 10^{18}$\,eV iron anti-nucleus is still confined in the Galactic plane for an extended time. On the right panel of Figure~\ref{Trajectories}, one can see that this anti-nucleus goes back and forth in the disk. It propagates especially along the regular field lines which are locally approximately oriented along the $x$ axis.

The right panel of Figure~\ref{Trajectories} shows the trajectories of 1, 4 and $10 \times 10^{19}$\,eV iron anti-nuclei. At $10^{19}$\,eV (red line), the CR is still strongly deflected before escaping the Galaxy. If one sums up all deflections along its trajectory, it exceeds $360^{\circ}$. This iron anti-nucleus is weakly deflected over distances up to $\sim 1$\,kpc. It is strongly deflected only locally, when it reaches regions with stronger turbulent magnetic field fluctuations. At $10^{20}$\,eV, the trajectories are fully ballistic, see the blue line. At such energies, one expects that iron nuclei suffer deflections of the order of $\sim 20^{\circ} - 40^{\circ}$ before escaping the Galaxy. The $4 \times 10^{19}$\,eV anti-nuclei are not diffusive any more. However, they still experience large deflections. For instance, there is a big wiggle on the 40\,EeV particle trajectory, see the green line in the right panel.

In principle, the best way to compute the anisotropy of Galactic CR at Earth is to use forward tracking. One should inject cosmic rays in the Galaxy at the source locations and only record the momenta of CR which cross a sphere around the Earth. The radius of this sphere should be small compared to the CR Larmor radius. This is, however, not feasible within reasonable computing times for the lowest rigidities we study. Therefore, we use a method first proposed in Ref.~\cite{Karakula:1972na}, and reused in more recent works such as Ref.~\cite{Lee:1995ki}. It consists in backtracking anti-particles with random initial momenta from the Earth to outside the Galaxy, and to record for each one the total path length in the source region. This corresponds to assuming a continuous and homogeneous source distribution inside the source region. Except for the Galactic center, potential Galactic CR sources should be transient. The method we use here assumes the existence of sufficiently frequent transient sources in the Galactic plane, so that the continuous source distribution hypothesis is fulfilled. For rare transient sources whose periodicity start to be comparable to the CR confinement time in the Galaxy, the current anisotropy may strongly differ from the average anisotropy~\cite{Pohl:2011ve}. Ref.~\cite{Pohl:2011ve} shows that if GRBs were to be the sources of Galactic CRs up to the ankle, strong variations of the Galactic CR flux and anisotropy should be expected on time scales of a hundred Myr. This would mean that CR anisotropy limits may be compatible with Pierre Auger measurements if we live in atypical times, when it drops below $\simeq2$\% at EeV energies~\cite{Abreu:2011ve}. Since one can hardly go beyond this statement for rare transients such as GRBs, we restrict our work to the case of sufficiently frequent transients. We will discuss below in more detail the domain of validity of this approximation, see Section~\ref{Spectrum}.

We count the length of particle trajectories contained in the source region without any weighting depending on the position inside the region. This corresponds to assuming that the sources are homogeneously distributed inside $-200\,{\rm pc} \leq z \leq 200\,{\rm pc}$ and $r \leq 20\,{\rm kpc}$. In practice, taking a more realistic source distribution with, for example, a source density decreasing with $|z|$, would only increase the Galactic CR anisotropy at Earth. The values presented below can be regarded as lower limits.

Finally, we compute the amplitude of the dipole of CR anisotropies and compare it to the upper limits presented by the Pierre Auger Collaboration, for energies $E \geq 1$\,EeV~\cite{Abreu:2011ve}. To do so, we associate to each of the $N$ backtracked cosmic rays a vector $\vec{v}(\theta,\phi)$ whose direction on the sky $(\theta,\phi)$ corresponds to the initial CR direction at Earth. Its length $|\vec{v}|$ corresponds to the trajectory path length in the source region. The dipole direction is given by the sum of all vectors $\sum_{i} {\vec v}_i$. In spherical coordinates $({\rm r},\theta,\phi)$, $\vec{v}=L\,(1+\mathcal{D} \cos \theta)\,\vec{u}_{r}$ for a dipole of amplitude $\mathcal{D}$ and direction $\vec{u}_{z}$, with $L = \frac{1}{N} \sum_i |\vec{v}_i|$ being the average length of vectors and $\theta$ the angle between $\vec{u}_{r}$ and $\vec{u}_{z}$. Since the vectors $\vec{v}$ are isotropically distributed,
\begin{displaymath}
 \sum_{i=1}^N {\vec v}_i = \int_{\theta=0}^{\pi} \int_{\phi=0}^{\pi} N \frac{\sin \theta d\theta d\phi}{4\pi}\, 2 |\vec{v}| \cos \theta \,{\vec u}_{\rm z} = \frac{NL\mathcal{D}}{3}\, {\vec u}_{\rm z}
\end{displaymath}
and the dipole strength equals
\begin{equation}
 \mathcal{D}=\frac{3}{NL} \left| \sum_{i=1}^N {\vec v}_i \right| \,.
\label{EqnD}
\end{equation}
If higher order multipoles are present, Eq.~(\ref{EqnD}) is unchanged because only the dipole has a non-zero contribution to the sum $\sum_i {\vec v}_i$.

\section{Anisotropy of Galactic Cosmic Rays predicted at Earth}
\label{Anisotropy}

In this Section, we compute the anisotropy at Earth of cosmic rays emitted by sources distributed in the Galactic plane, and compare it to the upper limits on the anisotropy as measured by the Pierre Auger Collaboration~\cite{Abreu:2011ve}. Within a given Galactic magnetic field model, if the predicted anisotropy of Galactic cosmic rays exceeds these limits at a given energy, a sufficiently large contribution of extragalactic CRs is required above that energy. Extragalactic CRs of energies below the GZK threshold around $\simeq4\times10^{19}\,$eV~\cite{gzk,stecker} tend to have anisotropies at the percent level or below because either a large number of sources at cosmological distances can contribute to the flux or, in case of relatively strong deflections in extragalactic magnetic fields, such deflections tend to wash out anisotropies over the long path lengths propagated over the typical energy loss distance. Resulting anisotropies of extragalactic cosmic rays can, therefore, easily be below the current upper limits. If sources at cosmological distances dominate the extragalactic CR flux, the motion of the Sun with respect to the CMB frame would induce an anisotropy of the extragalactic flux of $\simeq 0.6$\%~\cite{Kachelriess:2006aq}, due to the Compton-Getting effect~\cite{CGeffect}.

In subsection~\ref{Results_Anisotropy}, we present our predictions for the Galactic CR dipolar anisotropy and discuss its implications on the energy at which the transition from Galactic to extragalactic CR should occur. Figs.~\ref{Comp_200pc}--\ref{Dipole_1018} show computations for this anisotropy. Assuming in a first approximation that the extragalactic flux is isotropic, one can deduce from these figures the maximum contribution of Galactic CR sources to the total flux. For instance, at the transition from Galactic to extragalatic CRs, half of the CR flux is of Galactic origin and half of extragalactic origin. Therefore, the exact transition energy must be in the energy range where half of the dipole amplitude of Figs.~\ref{Comp_200pc}--\ref{Dipole_1018} does not overshoot the experimental upper limits on it. In~\ref{Anisotropy_1018protons}, we study in more detail the anisotropy predicted by Galactic protons below the ankle and its dependence on the GMF parameters.

\subsection{Dipole Amplitude and predicted Energy of the Transition}
\label{Results_Anisotropy}

We assume that the sources of Galactic CRs are located in a cylinder along the Galactic disc with height $z_{\max}$ and radius $r_{\max}$. Varying the extension of the source region from $r_{\max}=15$\,kpc to $r_{\max}=20$\,kpc, we verified that for rigidities $E/Z \lesssim 3 \times 10^{18}$\,eV, the dipole amplitude does not change by more than 3\%. For higher rigidities, the difference rarely exceed ($5-10$)\%. In practice, the conclusions below will not change for $r \leq 15$\,kpc or $r \leq 20$\,kpc. In the following we assume $r \leq 20$\,kpc.

Larger extensions of the source region along the $z-$direction would reduce the predicted anisotropy. We verified that, as long as the source region is less extended than $|z| \lesssim 500$\,pc, no strong modification of the results and conclusions would arise. We assume $|z| = 200$\,pc for most of the figures below.

For each of the following plots we backtrack $10^4$ particles. We find that this induces an error $\le \pm 3$\% on the predictions of the dipole amplitude. Reducing further this error to $\le \pm 1$\% could in principle be achieved by backtracking $\simeq 10$ times more CRs, but it is in practice impossible due to computing time reasons.

\begin{figure}
\begin{center}
\includegraphics[width=0.49\textwidth]{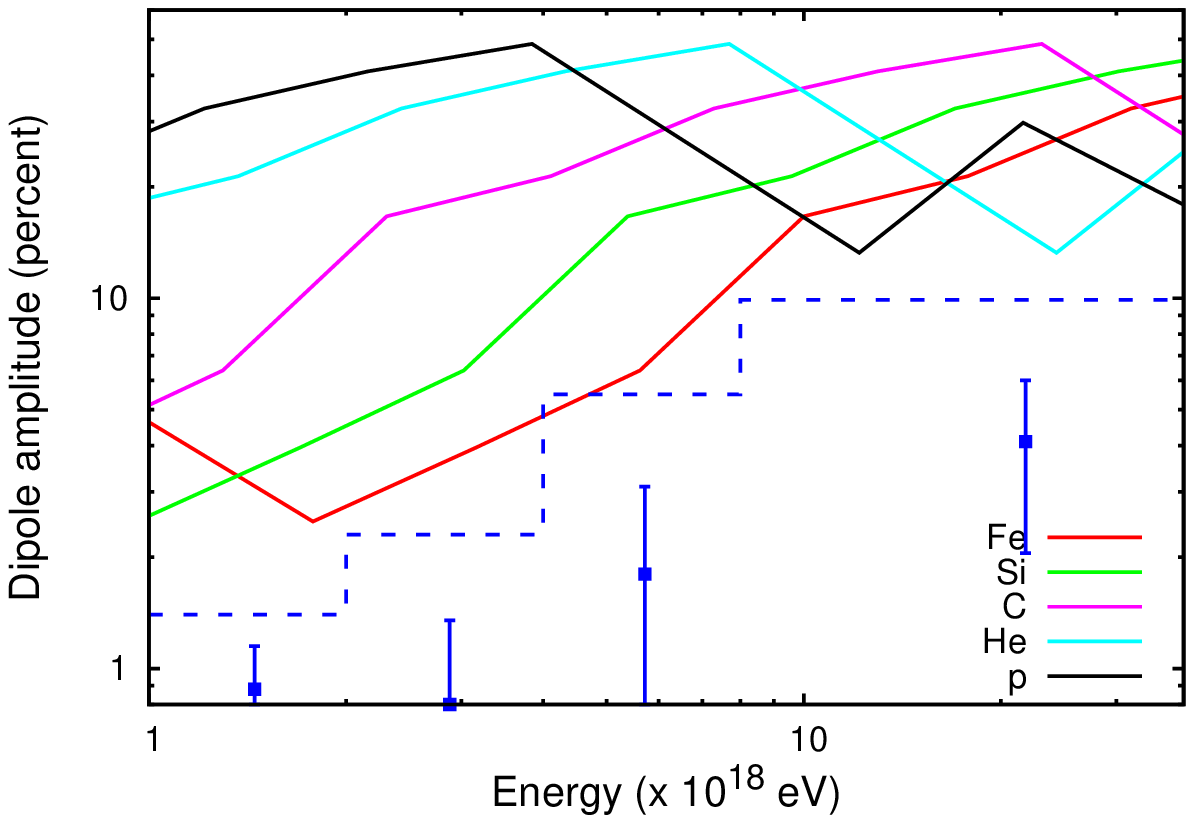}
\includegraphics[width=0.49\textwidth]{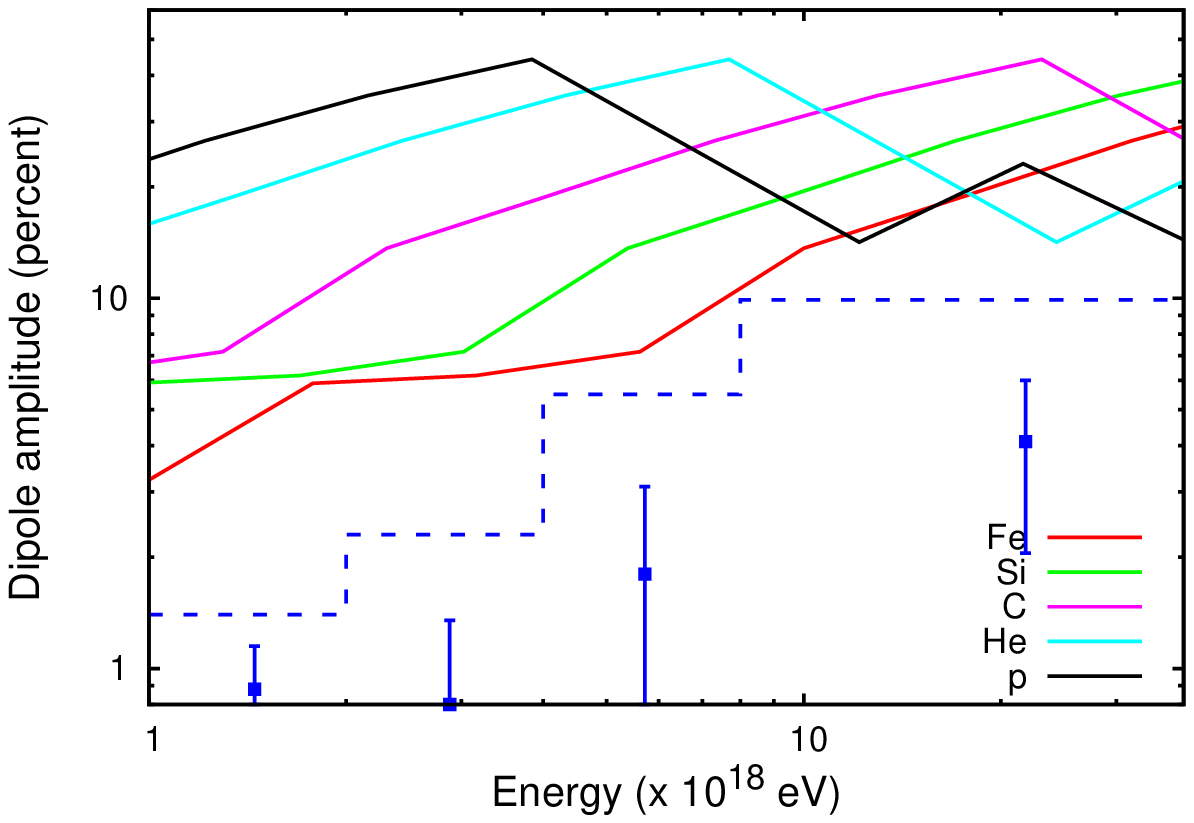}
\end{center}
\caption{\textbf{Left panel:} Predicted amplitude of the dipole as measured at Earth versus energy, for different primaries (p, He, C, Si, Fe) emitted by Galactic sources distributed in the region $-200\,{\rm pc} \leq z \leq 200\,{\rm pc}$ and $r \leq 20$\,kpc. The dashed blue line represents the 99\% C.L. upper limit on the dipole amplitude in right ascension as measured by the Pierre Auger Observatory~\cite{Abreu:2011ve}. Blue points represents the Pierre Auger measurements of the dipole in right ascension with the ``East-West method'' for the $1-2$\,EeV bin and with the ``Rayleigh analysis'' for the three other energy bins, according to Fig.~5 of Ref.~\cite{Abreu:2011ve}. The PTKN-BSS model was assumed for the regular GMF. The turbulent component is assumed to have a strength $B_0=4\,\mu$G, profile 1, and $z_0=2\,$kpc for its extension into the halo, with limiting length scales $L_{\min}=1\,$AU and $L_{\max}=200\,$pc; \textbf{Right panel:} Same as for the left panel, but for profile 2.}
\label{Comp_200pc}
\end{figure}

\begin{figure}
\begin{center}
\includegraphics[width=0.49\textwidth]{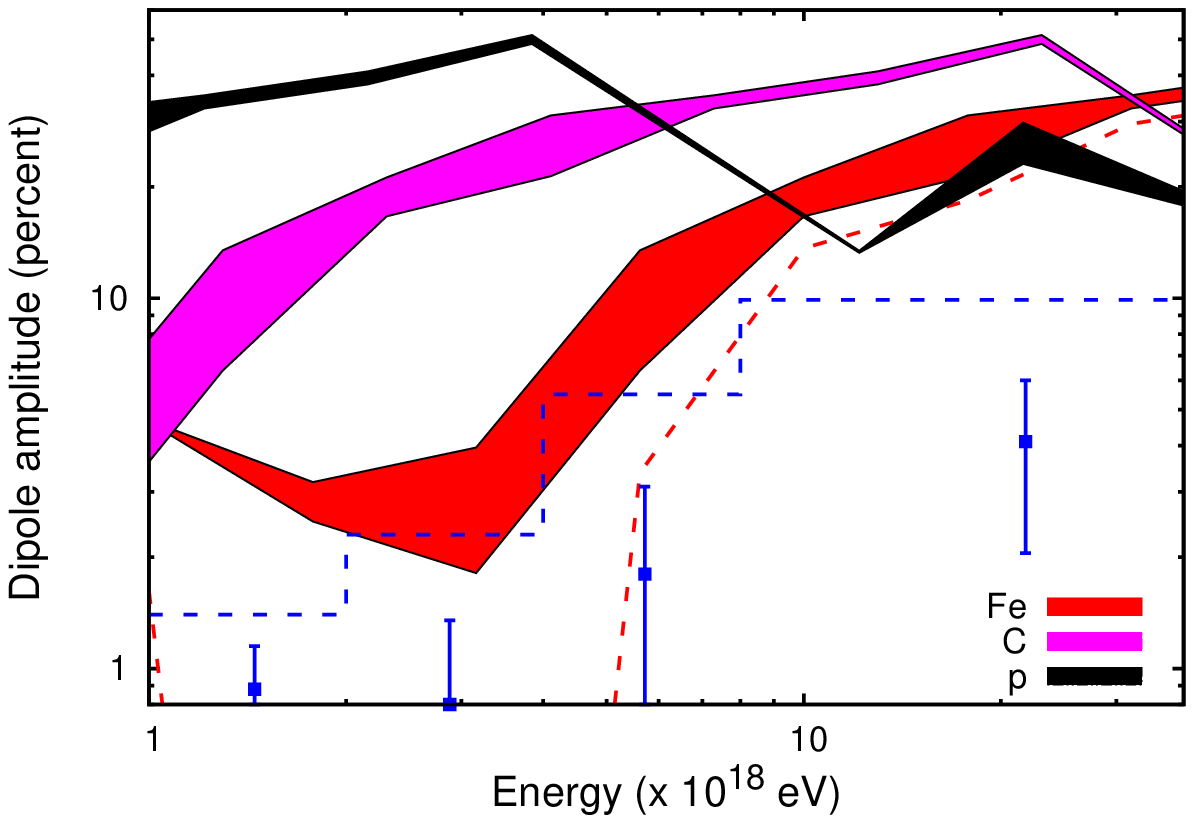}
\includegraphics[width=0.49\textwidth]{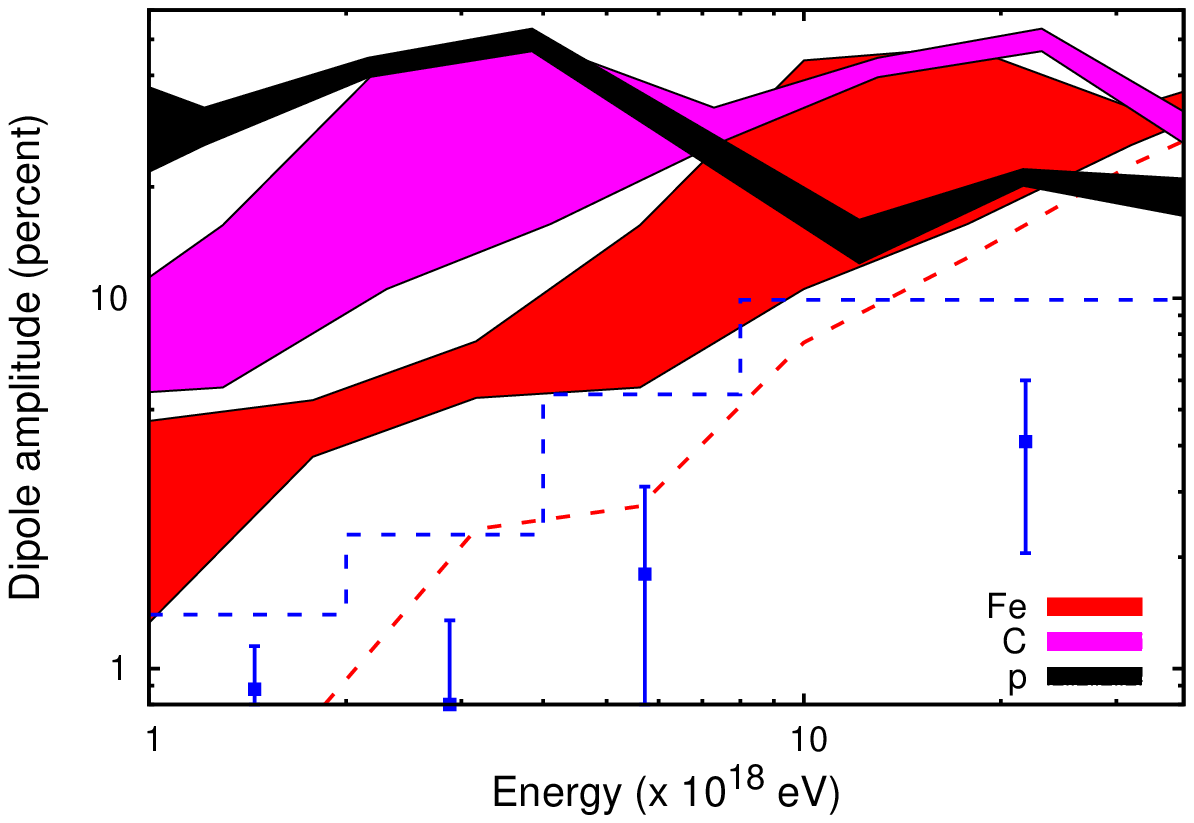}
\end{center}
\caption{Dependence of the predicted dipole amplitude on the turbulent field spectral index and on its maximum spatial variation scale. For comparison, Pierre Auger data~\cite{Abreu:2011ve} are shown in blue as in Fig.~\ref{Comp_200pc}. \textbf{Left panel:} Shaded area for $\alpha \in [3/2,5/3]$ (from Kraichnan to Kolmogorov); \textbf{Right panel:} Shaded area for $L_{\max} \in [100\,{\rm pc},300\,{\rm pc}]$. Red dashed lines for the lower error bars on the iron filled curve. For each plot, the values for all other parameters are set to those used in Fig.~\ref{Comp_200pc}.}
\label{Comp_Spectrum_Lmax}
\end{figure}

\begin{figure}
\begin{center}
\includegraphics[width=0.49\textwidth]{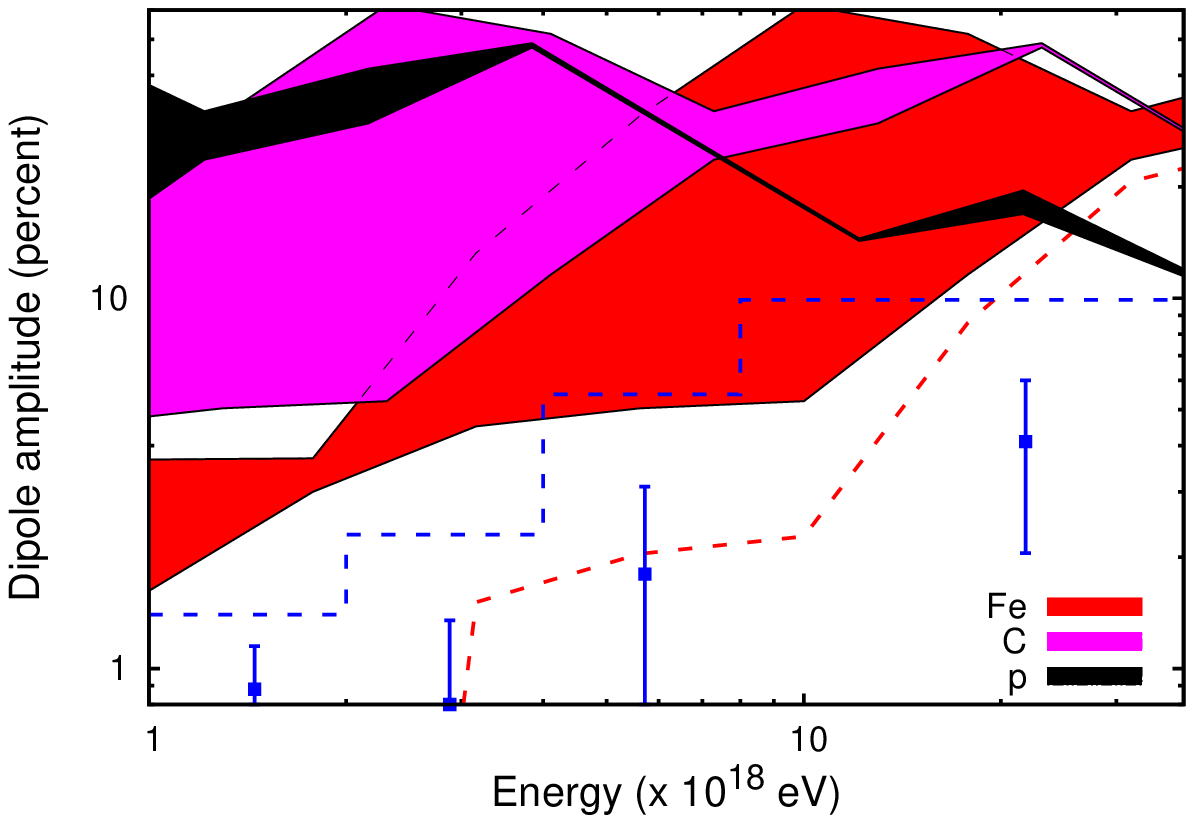}
\includegraphics[width=0.49\textwidth]{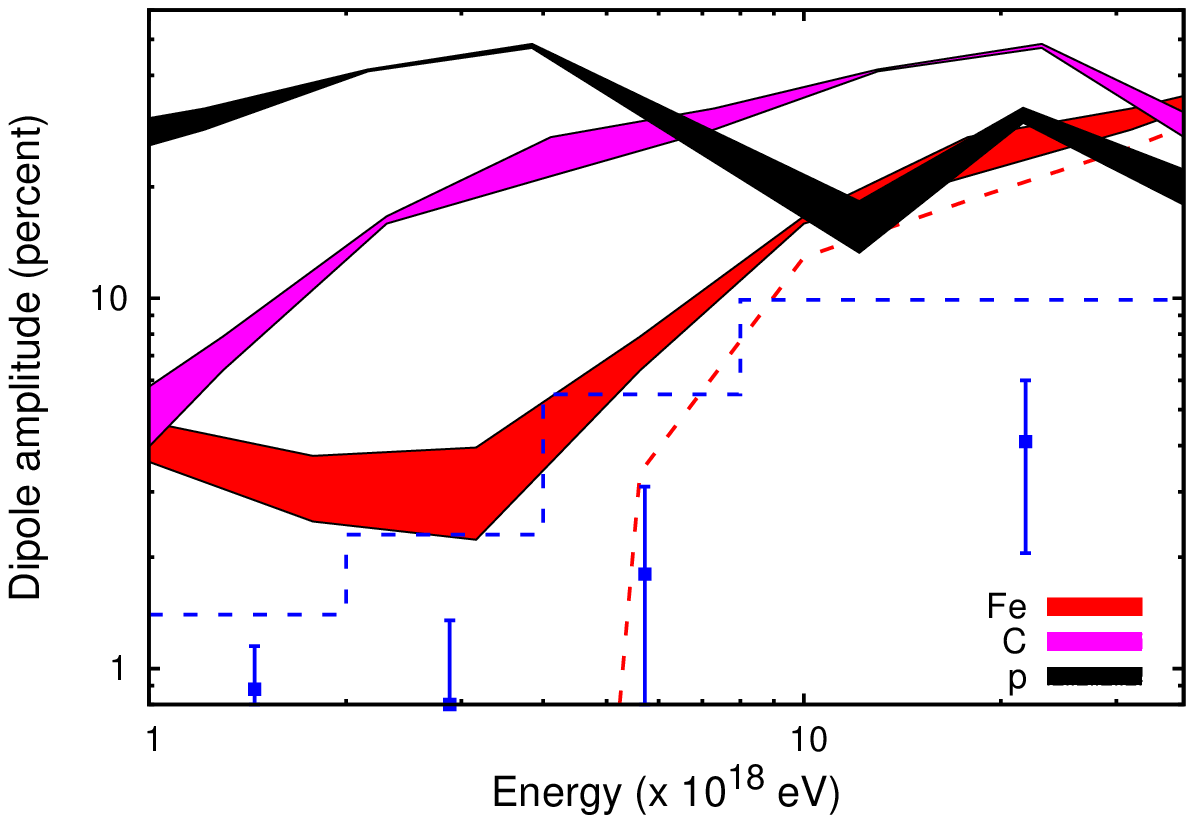}
\end{center}
\caption{Dependence of the predicted dipole amplitude on the turbulent field strength at Earth and on its extension in the Galactic halo. For comparison, Pierre Auger data~\cite{Abreu:2011ve} are shown in blue as in Fig.~\ref{Comp_200pc}. \textbf{Left panel:} Shaded area for $B_0 \in [2\,\mu{\rm G},8\,\mu{\rm G}]$; \textbf{Right panel:} Shaded area for $z_0 \in [2\,{\rm kpc},8\,{\rm kpc}]$. Red dashed lines for the lower error bars on the iron filled curve. For each plot, the values for all other parameters are set to those used in Fig.~\ref{Comp_200pc}.}
\label{Comp_B0_z0}
\end{figure}

Figure~\ref{Comp_200pc} presents the simulated predictions for the dipolar amplitude at Earth for several different Galactic CR primaries: protons, helium, carbon (representative for the CNO group), silicon and iron. CR sources are assumed to be located in the thin disk with $-200\,{\rm pc} \leq z \leq 200\,{\rm pc}$ and $r \leq 20$\,kpc. The PTKN-BSS model is assumed for the regular GMF, and the turbulent field parameters are taken to be $B_0=4\,\mu$G, $z_0=2\,$kpc, $L_{\min}=1\,$AU and $L_{\max}=200\,$pc. In the left panel, we use profile~1 given in Eq.~(\ref{profile1}) for the turbulent field, whereas in the right panel we take profile~2 from Eq.~(\ref{profile2}). We include in these figures measurements of, and 99\% C.L. upper limits on, the dipole amplitude in right ascension from the Pierre Auger Collaboration, as indicated in the captions. We assume that the true dipole vector does not lie in the equatorial plane, so that these upper limits in right ascension do not overconstrain significantly the total amplitude. In Fig.~\ref{Comp_200pc}, the predictions for the dipole amplitude for silicon and iron primaries may appear to exceed the Auger upper limits in the whole energy range we consider. However, since error bars on our computations are $\simeq \pm 3$\%, some parts of these silicon and iron lines are compatible with the Auger limits at low energies. For instance, one cannot exclude the dipole amplitude for 1--few\,EeV Galactic iron to be below the Auger limits. In the next two figures,~\ref{Comp_Spectrum_Lmax} and~\ref{Comp_B0_z0}, the red dashed lines represent the lower error bars on the red shaded bands. These bands represent the allowed ranges for the dipole amplitude with Galactic iron primaries, for different turbulent GMF parameters. In the energy ranges where this dashed line falls below Auger upper limits, a pure flux of Galactic iron CRs cannot be excluded on anisotropy grounds. In practice, we expect that with ten times larger statistics, predicted dipole amplitudes at the lowest rigidities ($E/Z \la {\rm few\,EeV}/26$), which are already smaller than a few percent, would fall further below the Auger limits. We have 10 sets of $10^4$ CRs for different turbulent GMF parameters. Their individual dipole directions look random at the lowest rigidities and if one adds up these ten sets, the resulting anisotropy for iron at $E = 10^{18}$\,eV falls to $\approx 0.6$\%, which is below the Auger limit. This gives an idea of what should be expected for ten times larger statistics. Therefore, below the ankle, both silicon and iron of Galactic origin are compatible with Pierre Auger Observatory limits. For the set of GMF parameters assumed here, in case of a predominantly heavy composition below the ankle and sufficiently frequent transient sources, CRs may still be of Galactic origin up to the ankle. Iron nuclei of Galactic origin up to $\simeq 10$\,EeV cannot currently be ruled out from the point of view of the CR anisotropy.

Depending on the composition at $E\simeq 10^{18}$\,eV, this has an important implication for the transition energy between Galactic and extragalactic CRs: For the set of GMF parameters assumed here, if the CR primary composition is predominantly light (p, He) or intermediate (C, N, O) at these energies, the predicted anisotropy at Earth would be larger than the 99\% C.L. upper limits from the Pierre Auger experiment if these nuclei were of Galactic origin, as seen in Fig.~\ref{Comp_200pc}. This implies that if the composition at $E\simeq 10^{18}$\,eV is measured to be light or intermediate, scenarios in which the transition from Galactic to extragalactic CRs occurs at the ankle are strongly disfavoured, at least for a wide range of GMF parameters. We investigate below the ranges of parameters for which this conclusion would be valid.

Figure~\ref{Comp_200pc} also shows that the conclusions above do not strongly depend on the turbulent field profile. For profile 2 (right panel), the predicted dipolar anisotropy grows slightly more slowly with energy than for profile 1 (left panel). This is expected because for profile 1 the gradient of the turbulent field tends to drive CRs towards larger $z$ in the Galactic halo slightly faster than for the constant field of profile 2. Since predicted anisotropies are not very different for the two profiles, we will mostly focus on profile 1 in the following.

We have also tested the dependence of these results on the regular GMF model. For rigidities $E/Z \gtrsim 3$\,EeV, the dipole amplitude and direction depends on the regular GMF model. However, the change of the dipole amplitude is too small to affect significantly our findings. Moreover, the PTKN-BSS model which we use in all the following figures is one of the models with the lowest dipole amplitudes among those tested.

Figures~\ref{Comp_Spectrum_Lmax} and~\ref{Comp_B0_z0} present the dependence of the previous results on the other turbulent GMF parameters, which are poorly constrained: the index $\alpha$ of the fluctuation spectrum (Fig.~\ref{Comp_Spectrum_Lmax} - left panel), the maximal length of field fluctuations $L_{\max}$ (Fig.~\ref{Comp_Spectrum_Lmax}, right panel), the field strength normalization $B_0$ (Fig.~\ref{Comp_B0_z0}, left panel) and the scale height $z_0$ (Fig.~\ref{Comp_B0_z0} - right panel). The shaded areas of the filled cures represent the relative change of the results for p, C and Fe primaries when varying separately the four above parameters and keeping all other parameters at the values in Fig.~\ref{Comp_200pc}. The red dashed line is computed as the lower boundary of shaded areas for iron primaries minus 3\% from our statistical uncertainties.

At low rigidities $E/Z \lesssim 4$\,EeV, the dipole amplitude grows with $E/Z$ as expected. At larger rigidities, CRs start to enter the ballistic regime and higher order multipoles start to make a significant contribution to the total Galactic CR anisotropy at Earth. As seen in Figs.~\ref{Comp_Spectrum_Lmax} and~\ref{Comp_B0_z0}, the dipole amplitude may then become smaller and/or vary with $E/Z$.
Thus a decrease of the {\em dipole\/} amplitude at high energies does not necessarily imply that the distribution of CR arrival directions at Earth becomes more isotropic.

The widths of the filled curves indicate that results mostly vary with the turbulent GMF strength and its maximum spatial variation scales. Results are less sensitive to the spectral index of the field and no strong difference in the dipole amplitude is found between the fields with Kolmogorov and Kraichnan spectra. The amplitudes at $E/Z \gtrsim 3$\,EeV/26 are slightly larger for the Kraichnan spectrum because less power is present in the large length scale modes, which are relevant at such rigidities, than for the Kolmogorov spectrum. Results are only marginally sensitive to the extension $z_0$ of the turbulent field into the halo, see right panel of Fig.~\ref{Comp_B0_z0}. This is due to the fact that CRs which escape the source region and propagate to large $z$ in the halo rarely come back to the source region.

Results are mostly sensitive to $L_{\max}$ and $B_0$. The maximal length scale of the turbulence $L_{\max}$ determines up to which energy CRs still scatter on the turbulent field inhomogeneities. For larger $L_{\max}$, CRs can be diffusive up to larger energies, which therefore reduces their expected anisotropy at Earth. One can see in the right panel of Figure~\ref{Comp_Spectrum_Lmax} that for $L_{\max}=300$\,pc and $B_0=4\,\mu$G, the dipole amplitude below $E \simeq 15$\,EeV for a pure iron composition may be compatible with the current 99\% C.L. upper limits from the Pierre Auger Observatory.

The dependence of our results on the turbulent field strength $B_0$ is very strong, see the left panel of Fig.~\ref{Comp_B0_z0}. The upper parts of the shaded areas correspond to $B_0=2\,\mu$G and the lower to $B_0=8\,\mu$G. For $B_0=8\,\mu$G, the anisotropy at Earth of iron primaries is \textit{a priori} compatible with the Pierre Auger upper limits up to $E\simeq20$\,EeV. For $B_0=2\,\mu$G, the dipole amplitude starts to overshoot the Pierre Auger limits around $E\simeq 3\times 10^{18}$\,eV, while for $B_0=4\,\mu$G, the amplitude starts to exceed the Pierre Auger limits around 10\,EeV.
For all cases, a light or intermediate composition at $E \simeq 10^{18}$\,eV would exceed the Pierre Auger upper limits and rule out the ankle as the transition from Galactic to extragalactic cosmic rays.

\subsection{Dipole Amplitude at $E/Z=10^{18}$\,eV}
\label{Anisotropy_1018protons}

In this section we demonstrate that for any reasonable combination of turbulent GMF parameters, at $E\sim10^{18}$\,eV the dipolar anisotropy predicted by light primaries of Galactic origin is always larger than the observational limit from the Pierre Auger experiment. Therefore, having a reliable composition measurements at such energies is crucial for knowing if the ankle can or cannot be the signature of the transition from Galactic to extragalactic CRs. Low energy extensions such as HEAT~\cite{HEAT} and AMIGA~\cite{AMIGA} can solve this important question.

\begin{figure}
\begin{center}
\includegraphics[width=0.49\textwidth]{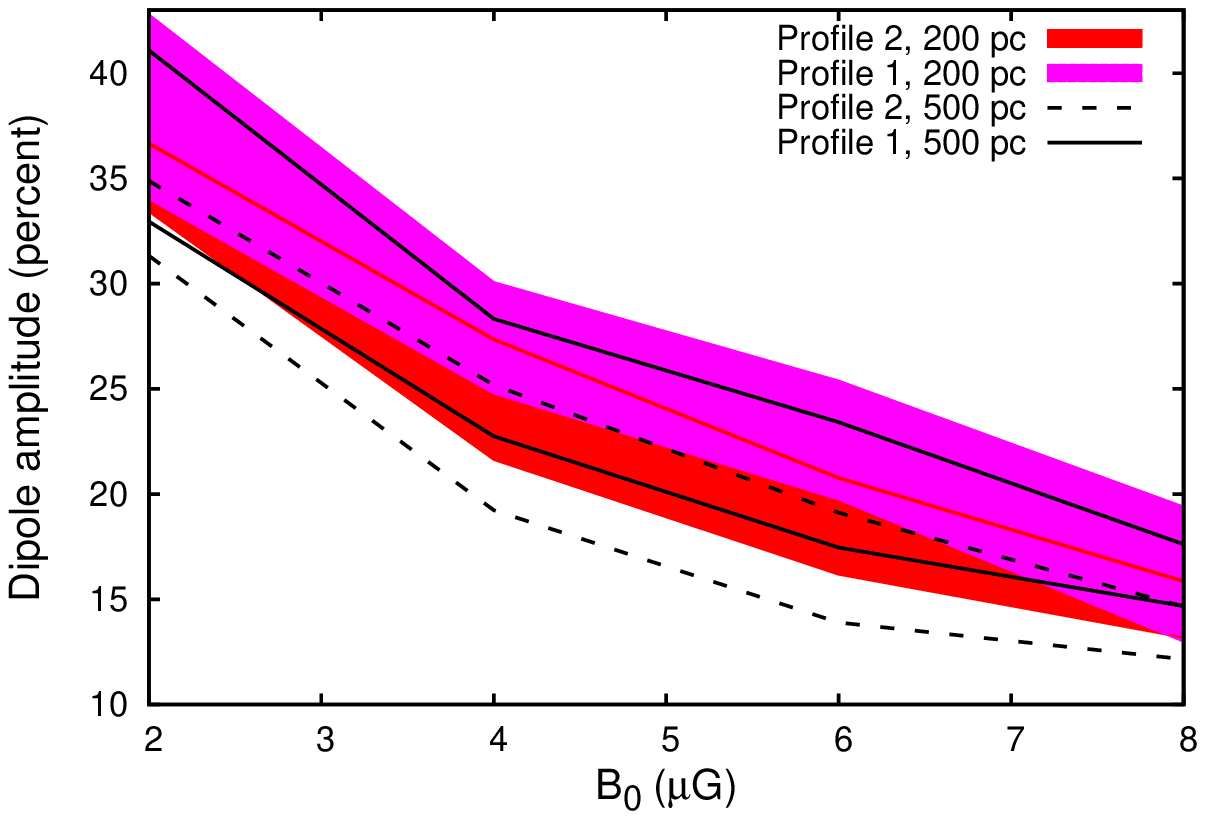}
\includegraphics[width=0.49\textwidth]{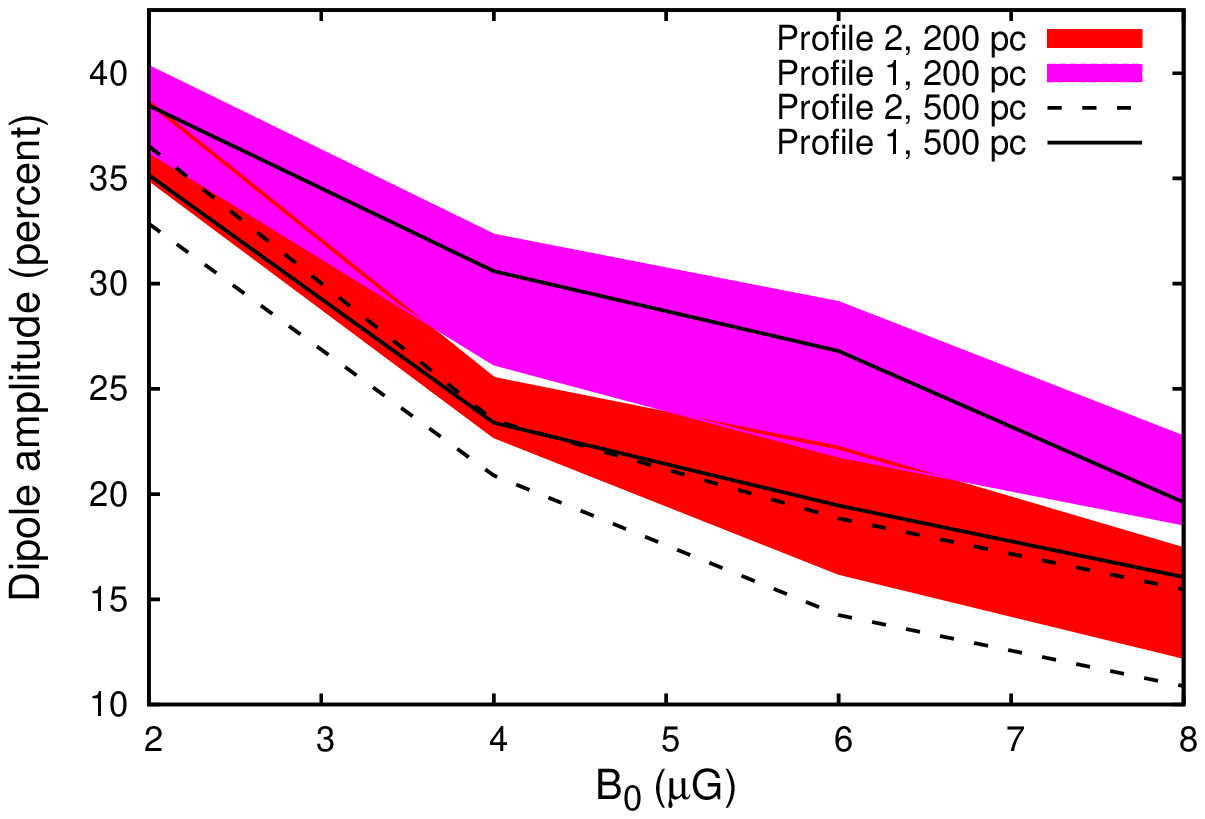}
\end{center}
\caption{Predicted amplitude of the dipole versus the turbulent Galactic magnetic field strength, for $E/Z=10^{18}$\,eV cosmic rays emitted by Galactic sources distributed in the region with $r \leq 20\,{\rm kpc}$, and $-200\,{\rm pc} \leq z \leq 200\,{\rm pc}$ or $-500\,{\rm pc} \leq z \leq 500\,{\rm pc}$, respectively, as indicated. Profiles 1 and 2 (see subsection~\ref{Method_Anisotropy}) for the turbulent Galactic magnetic field profile along $z$, as indicated. For the regular GMF the PTKN-BSS model is assumed. Shaded or delimited areas correspond to $z_0$ varying in the range $1-8$\,kpc. For the turbulent component a Kolmogorov \textbf{(left panel)} or Kraichnan \textbf{(right panel)} spectrum with $L_{\min}=1\,$AU and $L_{\max}=200\,$pc is assumed.}
\label{Dipole_1018}
\end{figure}

Figure~\ref{Dipole_1018} shows how the strength predicted at Earth of the dipole amplitude of 1\,EeV protons from Galactic sources depends on the turbulent field rms strength $B_0=2-8\,\mu$G when the scale height $z_0$ is allowed to vary in the range 1 to 8\,kpc (shaded areas). Both turbulent field profiles are tested, and a $\pm 500$\,pc width source region is also tested. Both for Kolmogorov (left panel) and Kraichnan spectra (right panel), predicted dipole amplitudes are above 10\%, considerably higher than the $\sim2$\% upper limit from the Pierre Auger experiment at such energies.
Therefore, the ankle cannot be the signature of the transition from Galactic to extragalactic CRs, if the contribution of Galactic protons to the CR flux at 1\,EeV is larger than $\simeq20$\%. Such a scenario would be consistent with the dip model~\cite{dip}. A CR flux at 1\,EeV consisting mostly of light nuclei cannot be of Galactic origin, except in the very unlikely case of $B_0 \gg 10\,\mu$G.

\section{Energy Spectrum of Galactic Cosmic Rays and Sources contributing at Earth}
\label{Spectrum}

\begin{figure}
\begin{center}
\includegraphics[width=0.49\textwidth]{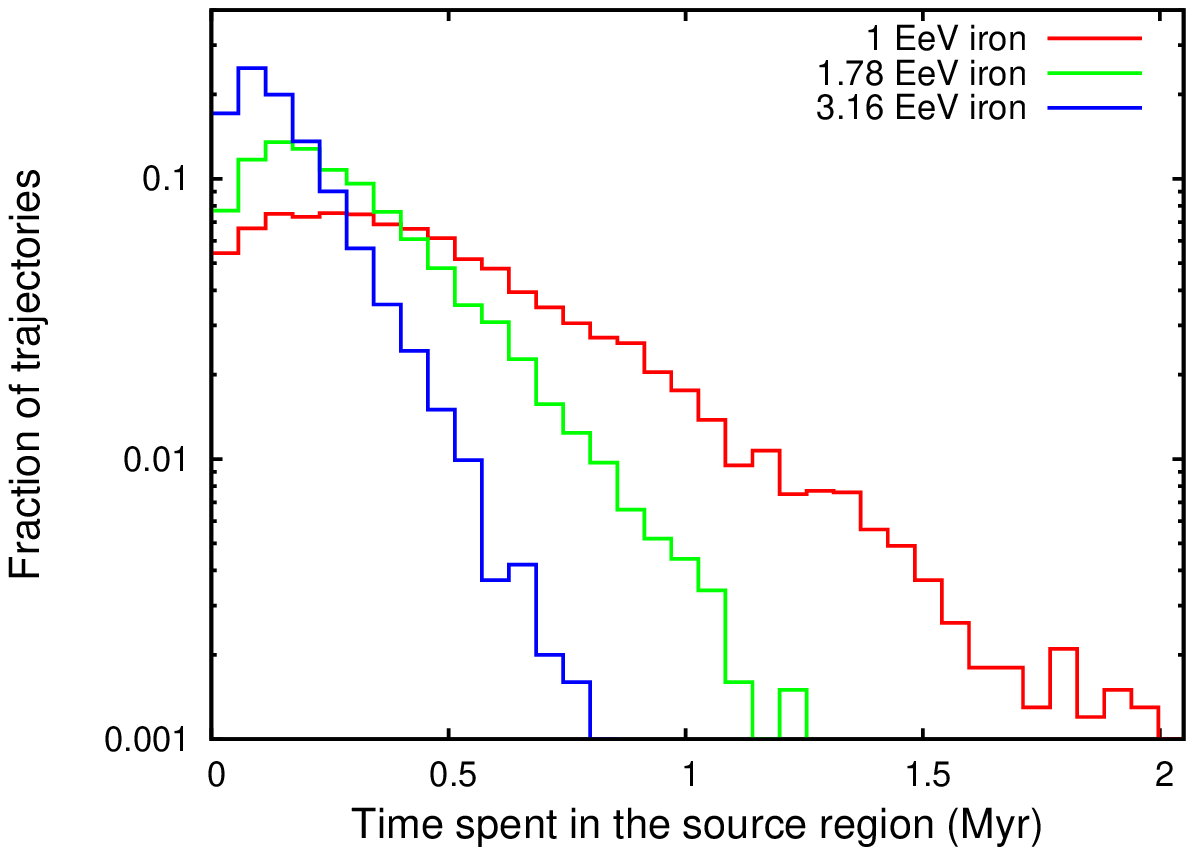}
\includegraphics[width=0.49\textwidth]{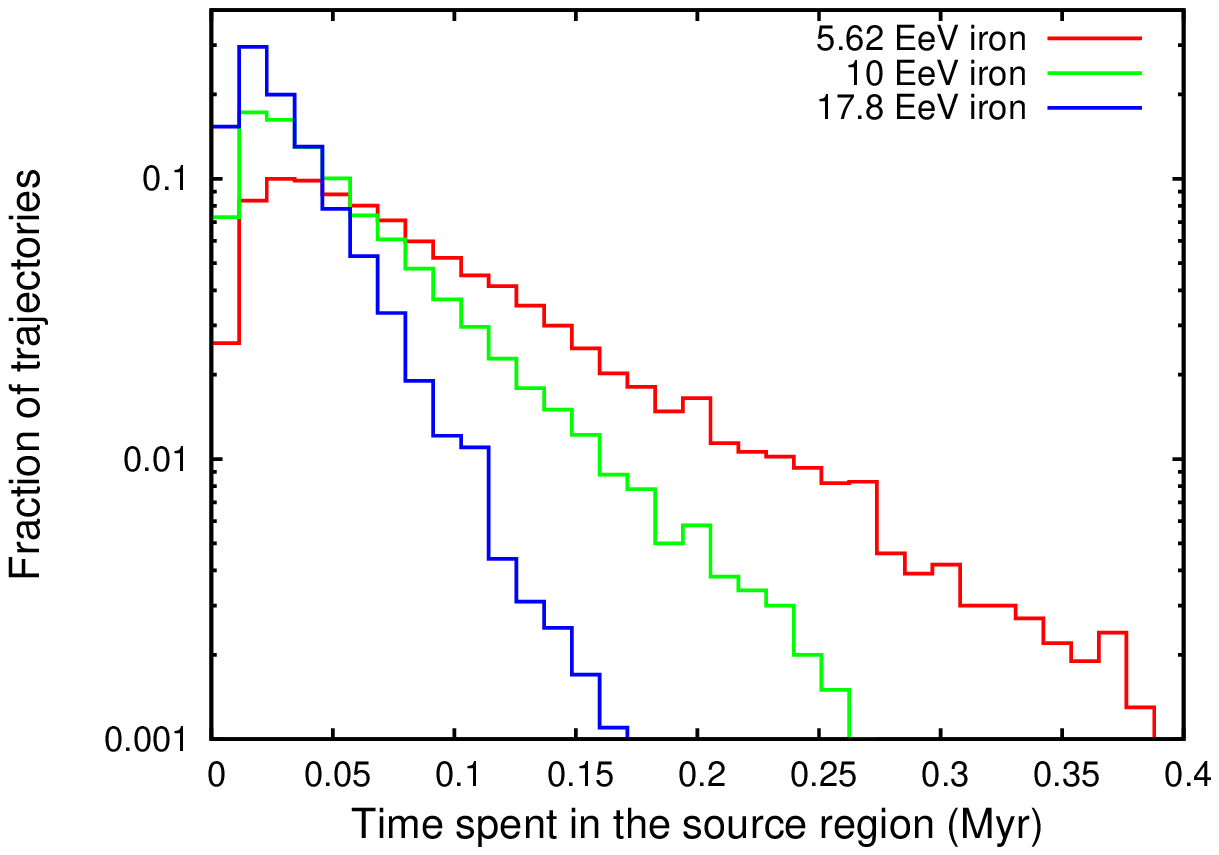}
\end{center}
\caption{Histograms of fractions of cosmic rays spending a given time in the source region ($-200\,{\rm pc} \leq z \leq 200\,{\rm pc}$ and $r \leq 20\,{\rm kpc}$), for rigidities $E/Z = (1,\,1.78,\,3.16) \times10^{18}\,$eV/26 \textbf{(left panel)}, and $E/Z = (5.62,\,10,\,17.8) \times10^{18}\,$eV/26 \textbf{(right panel)}. For the regular GMF the PTKN-BSS model is assumed. For the turbulent component a Kolmogorov spectrum with $L_{\min}=1\,$AU and $L_{\max}=200\,$pc, strength $B_0=4\,\mu$G and scale height $z_0=2\,$kpc is assumed.}
\label{EscapeLength}
\end{figure}

Figure~\ref{EscapeLength} presents histograms of the relative fraction of CRs spending a certain time in the source region, for rigidities ranging from 1\,EeV/26 to 17.8\,EeV/26. We use the PTKN-BSS model for the regular GMF component. The turbulent GMF strength is set to $B_0=4\,\mu$G, its extension in the halo to $z_0=2\,$kpc. We take a Kolmogorov spectrum with maximum spatial variation scale $L_{\max}=200\,$pc.

With such parameters, the average time spent in the source region for cosmic rays with $E/Z=1$\,EeV/26 is $\simeq0.5$\,Myr. The CR escape times from the magnetized region of the Galaxy defined as $-10\,{\rm kpc}\lesssim z \lesssim 10\,{\rm kpc}$ and $r \lesssim 20\,{\rm kpc}$, are found to be $\simeq 5$ times larger than the times spent in the source region for the turbulent GMF profile 1. CRs which escape the Galactic thin disk containing CR sources still stay a non-negligible amount of time in the halo compared to the time spent in the source region.

For 1\,EeV iron nuclei (red curve in Fig.~\ref{EscapeLength} - left panel), only a few percent of CRs stay more than 1\,Myr in the source region. This implies that rare transient sources such as gamma ray bursts (GRBs) are very unlikely to be sources of Galactic CRs in the sub-ankle region, even if heavy nuclei were able to escape such sources.

For 10\,EeV iron nuclei, the average time spent in the source region is $\simeq0.06$\,Myr, which is nearly ten times smaller than at 1\,EeV. We found that the average time spent in the source region is approximately proportional to $1/E$ -or slightly softer- in the rigidity range $E/Z \in [1\,{\rm EeV}/26,\,20\,{\rm EeV}/26]$. This implies that for a source injection spectrum proportional to $E^{\alpha}$, the spectrum at Earth approximately goes as $E^{\alpha - 1}$. Interestingly, this result is compatible with predictions from diffusion in the regime where the Larmor radius is larger than the coherence length of the turbulent field: The distance traveled by CRs within a given amount of time is proportional to the square root of the diffusion coefficient, and the diffusion coefficient is proportional to $E^2$ when Larmor radii are larger than $\approx L_{\rm c}$~\cite{Casse:2001be}. We found no significant change to this conclusion when varying turbulent GMF parameters in the ranges tested in the previous section. However, the time spent in the source region for $B_0=8\,\mu$G is $\simeq 30$\% smaller than for $B_0=4\,\mu$G at $E/Z\sim10^{18}$\,eV. For stronger turbulent fields, the larger turbulent field gradient towards $z$ tends to make CRs leave the source region faster.

\begin{figure}
\begin{center}
\includegraphics[width=0.49\textwidth]{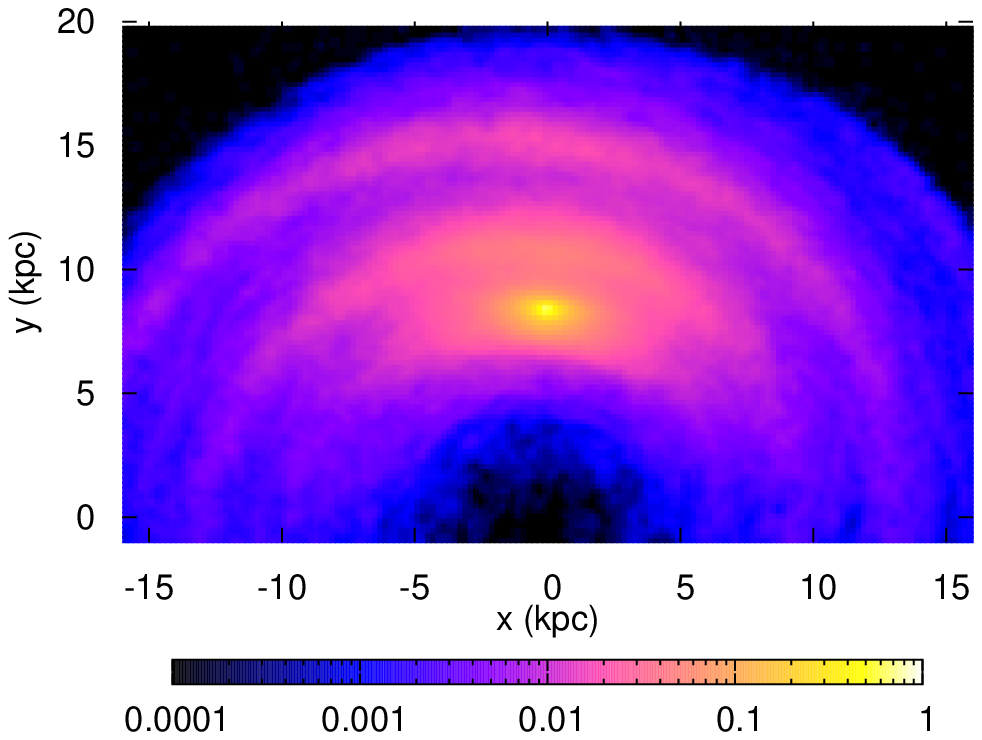}
\includegraphics[width=0.49\textwidth]{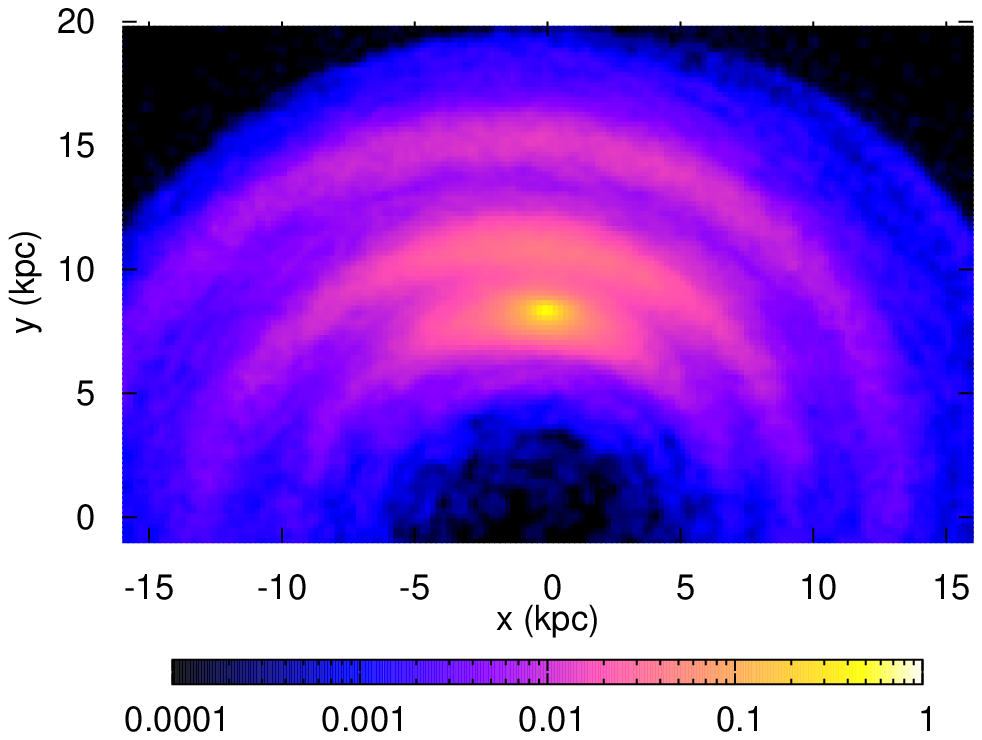}
\includegraphics[width=0.49\textwidth]{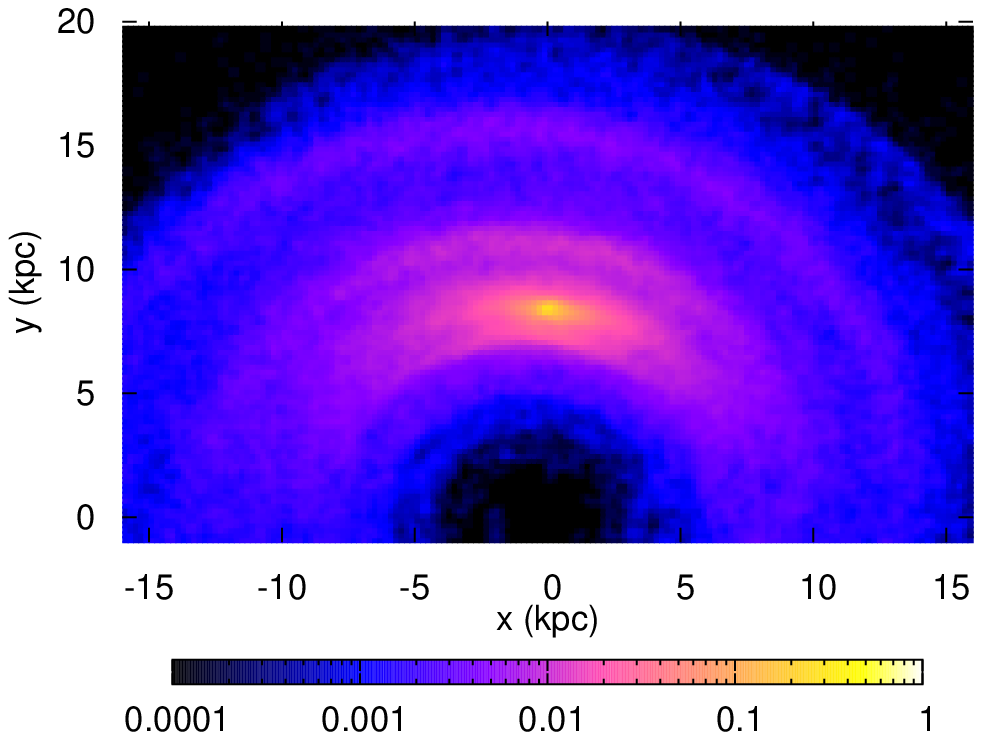}
\includegraphics[width=0.49\textwidth]{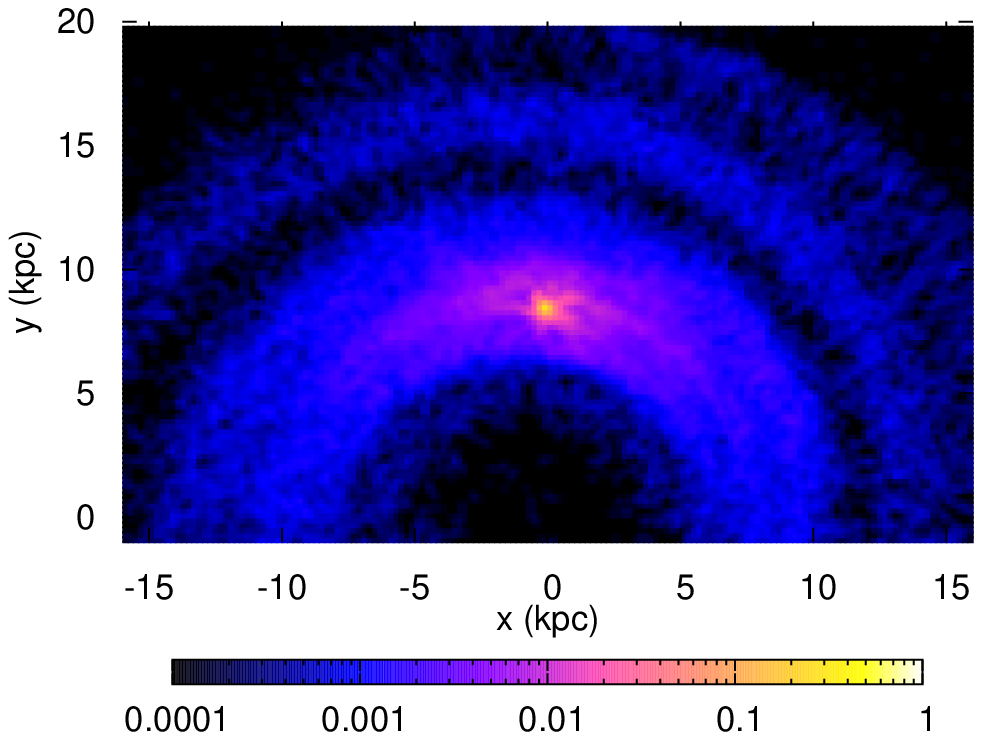}
\end{center}
\caption{Relative contributions per volume to the Galactic CR flux observed at Earth: Fraction of all particles backtraced from the Earth which cross (200\,pc)$^3$ cubes located in the source region. Earth is located at $(x,y)=(0,8.5\,{\rm kpc})$. Galactic plane in the plane of the panels. \textbf{Upper left panel:} For rigidity $E/Z=10^{18}$\,eV/26; \textbf{Upper right panel:} $E/Z=3 \times 10^{18}$\,eV/26; \textbf{Lower left panel:} $E/Z=10^{19}$\,eV/26; \textbf{Lower right panel:} $E/Z=3 \times 10^{19}$\,eV/26. Same Galactic magnetic field parameters as in Fig.~\ref{EscapeLength}.}
\label{GalContributions}
\end{figure}

We show in Fig.~\ref{GalContributions}, which regions of the Galactic plane are passed most by CRs backtraced from the Earth for $E/Z=10^{18}$\,eV/26 (upper left panel), $3\times 10^{18}$\,eV/26 (upper right), $10^{19}$\,eV/26 (lower left) and $3\times 10^{19}$\,eV/26 (lower right). This equivalently shows which parts of the source region contribute most to the Galactic CR flux detected at Earth. The color code presents the fraction of all particles backtraced from the Earth which pass cubes of 200\,pc lateral size located in the Galactic disk. Here, CRs are not counted more than once. Location of Earth is marked by the bright spot at $(x,y)=(0,8.5\,{\rm kpc})$. One can clearly see the shapes of the spiral arms present in the model of the regular GMF model. CRs indeed diffuse or propagate faster along the regular field direction. The Galactic center region appears strongly demagnified, which makes it unlikely to significantly contribute to the observed fluxes. This is due to the stronger field in the disk and especially towards the Galactic bulge. For the forward tracking point of view, CRs potentially emitted by the Galactic center would escape from the Galactic thin disk which contains the Earth, and propagate towards larger $z$, before reaching Earth. At $E/Z=3 \times 10^{19}$\,eV/26 (lower right panel), the CRs start to be in the ballistic regime and only the region within a few hundreds of parsecs from Earth could contain sources. At these rigidities, CRs cannot have Galactic origin any more because anisotropies would exceed the Pierre Auger upper limits on the dipole anisotropy, as discussed in the previous section.

When the confinement time of CRs in the source region ($\simeq 0.5$\,Myr for 1\,EeV iron nuclei) starts to be comparable to the period between two potential Galactic UHECR sources, the continuous source distribution approximation used in this paper breaks down. For GRBs, this happens at $E/Z \gtrsim (0.1 - 1)$\,EeV/26. In this case, the expected Galactic CR flux at Earth and its anisotropy would strongly vary on time scales of several Myr, see for instance Ref.~\cite{Pohl:2011ve}. The CR anisotropy at Earth may then be substantially smaller or larger than those computed for a continuous source distribution. Potential sources such as magnetars are expected to have a larger rate of $\sim 1$ per 1000 years. For such rates, $\sim (0.5\,{\rm Myr})/(1000\,{\rm yr})\sim$~several hundreds of sources would contribute to the flux observed at $E/Z=10^{18}$\,eV/26. This larger number of sources substantially reduces the fluctuations in time of the Galactic CR flux and anisotropy at Earth. Therefore, in this case, the anisotropy does not significantly differ from the values presented in this work, and the continuous source distribution approximation is valid.

\begin{figure}
\begin{center}
\includegraphics[width=0.49\textwidth]{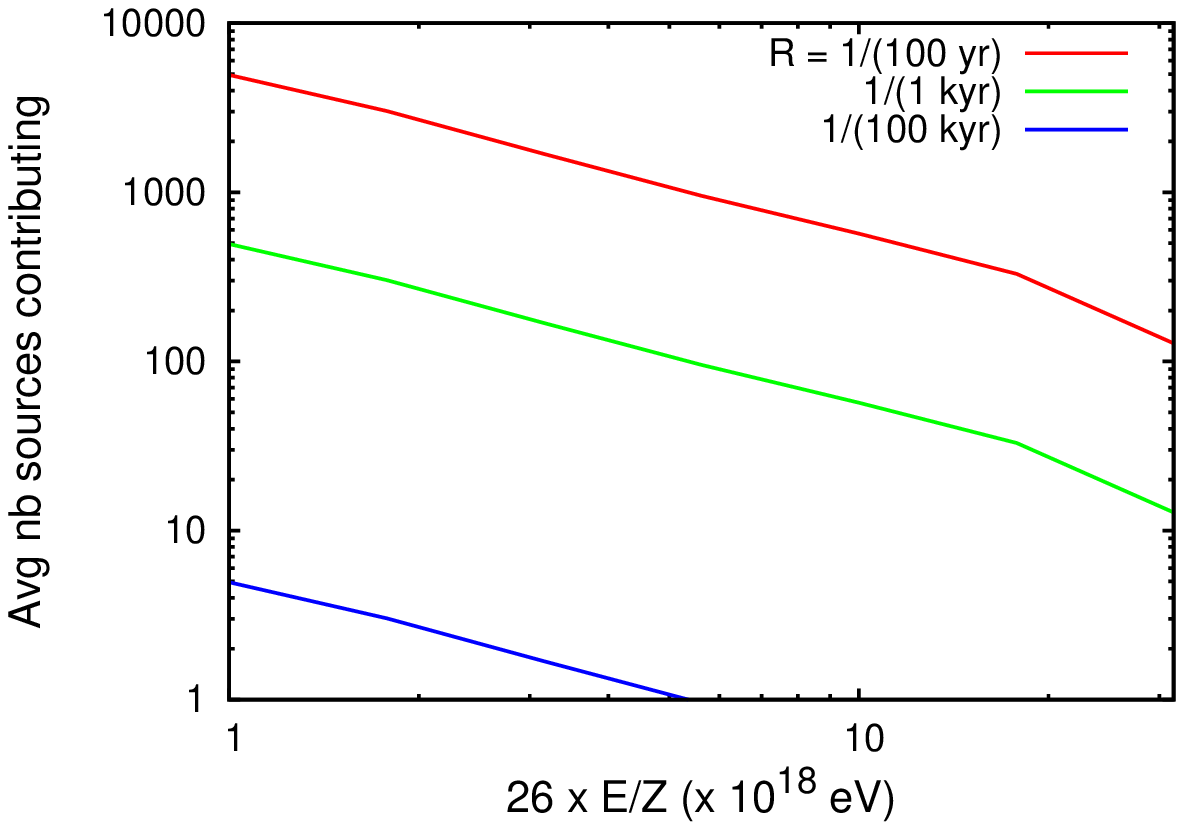}
\includegraphics[width=0.49\textwidth]{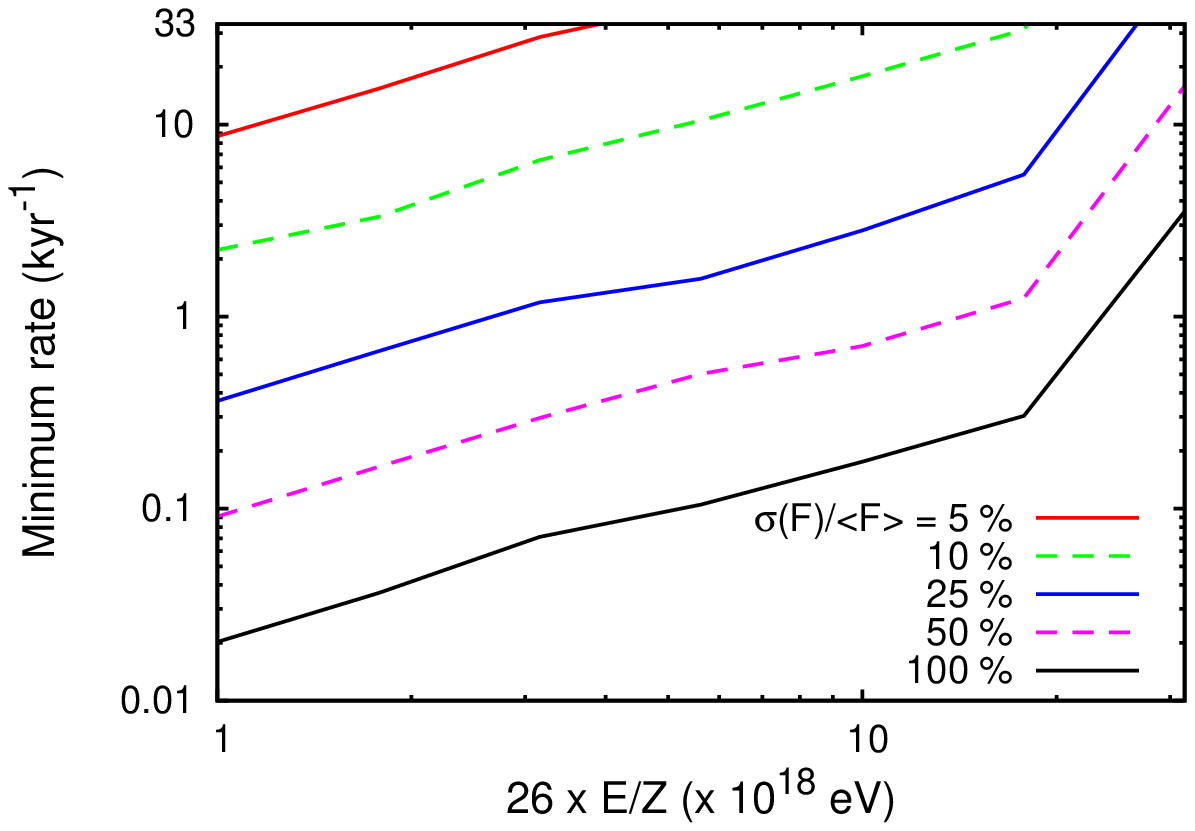}
\end{center}
\caption{\textbf{Left panel:} Estimate for the number of sources that would contribute to the Galactic CR flux at Earth versus $26 \times E/Z$, for three different source rates $\mathcal{R}=1/(100\,{\rm yr}),\,1/(1\,{\rm kyr}),\,1/(100\,{\rm kyr})$; \textbf{Right panel:} Estimate for the minimum rate $\mathcal{R}$ of Galactic CR sources that would be required to maintain relative fluctuations of the Galactic CR flux at Earth $\sigma (F)/\left\langle F \right\rangle$ below $\approx 5$, 10, 25, 50, 100\%, versus $26 \times E/Z$. For both panels, same Galactic magnetic field parameters as in Fig.~\ref{EscapeLength}.}
\label{Variability}
\end{figure}

Let us now estimate more quantitatively when the continuous source distribution approximation breaks down. The average number of sources that would contribute to the Galactic CR flux at one given rigidity can be estimated as the average time spent by CRs in the source region multiplied by the source rate $\mathcal{R}$. We plot in Fig.~\ref{Variability} (left panel) this estimate of the number of contributing sources, for three different rates $\mathcal{R}=10,\,1,\,0.01\,{\rm kyr}^{-1}$ and for CR rigidities in the range $E/Z=(1-32)\,{\rm EeV}/26$. We assume here for the turbulent GMF a Kolmogorov spectrum with $L_{\max}=200\,$pc, strength $B_0=4\,\mu$G and scale height $z_0=2\,$kpc. For sources with a rate $\mathcal{R}=1\,{\rm kyr}^{-1}$ (green line) similar to that expected for magnetars, the average number of contributing sources stays above $\gtrsim100$, up to the ankle for iron nuclei. It decreases from $\sim500$ for 1\,EeV iron to only $\sim10$ at 32\,EeV. As shown below, $\sim10$ sources is too small for the continuous source distribution approximation to be valid. With a ten times larger source rate $\mathcal{R}=10\,{\rm kyr}^{-1}$ (red line), there would be $\gtrsim100$ sources contributing for all the explored rigidity range, but $10\,{\rm kyr}^{-1}$ is of the order of the Galactic supernovae rate which looks very unlikely for extreme CR accelerators. The blue line corresponds to rarer transients $\mathcal{R}=0.01\,{\rm kyr}^{-1}$. At the ankle, only one source would contribute in average. Sources such as Galactic GRBs with $\mathcal{R}\sim1\,{\rm Myr}^{-1}$ cannot be described by the continuous source distribution approximation. However, such sources are unlikely to be responsible for the sub-ankle CR flux if it were to be of Galactic origin. Indeed, one can for example hardly match the bumpy CR spectrum resulting from rare Galactic transients to the observed smooth power law spectrum~\cite{Pohl:2011ve}.

We provide in Fig.~\ref{Variability} (right panel) an estimate of the minimum rate $\mathcal{R}$ of Galactic sources that would be required to maintain relative fluctuations of the flux $\sigma (F)/\left\langle F \right\rangle$ below five given thresholds ($\approx 5$, 10, 25, 50, 100\%) on $\gg \mathcal{R}^{-1}$ time scales. Since we follow individual CR trajectories, we cannot directly provide $\sigma (F)/\left\langle F \right\rangle$ for computing time reasons. We can however estimate its value: The panels of Fig.~\ref{GalContributions} also give an estimate of the total flux that would be received in any point of the Galactic disk from one source located at $(x,y)=(0,8.5\,{\rm kpc})$. Looking for the total flux received at the Earth position from $N$ sources -with same power- located in $N$ random positions in the Galactic disk is then roughly equivalent to putting $N$ observers in $N$ random locations in the disk and summing up the total flux they receive from the single source located at $(x,y)=(0,8.5\,{\rm kpc})$. We compute the flux $F$ in the latter way for $10^{4}$ different configurations, using the computations of Fig.~\ref{Variability} (left panel) for $N$. This yields the estimate for $\sigma (F)/\left\langle F \right\rangle$ that is used in Fig.~\ref{Variability} (right panel). To maintain $\sigma (F)/\left\langle F \right\rangle$ below $\approx 5$\% (red solid line) at the ankle for Galactic iron primaries, sources with rates comparable with that of Galactic supernovae $\mathcal{R} \approx 30$\,kyr$^{-1}$ would be needed. For rates $\mathcal{R} \sim 1$\,kyr$^{-1}$, $\sigma (F)/\left\langle F \right\rangle \approx 10$\% for 1\,EeV iron nuclei and remains $\lesssim 25$\% below the ankle. In this case, the continuous source distribution approximation is still valid. For energies $E\gtrsim (10 - 20)$\,EeV, it quickly starts to break down: For 20\,EeV (resp. 30\,EeV) iron nuclei, $\sigma (F)/\left\langle F \right\rangle \approx 50$\% (resp. $\approx 100$\%). For Galactic source rates $\mathcal{R} \lesssim 0.01$\,kyr$^{-1}$, $\sigma (F)/\left\langle F \right\rangle$ exceeds 100\% above $\sim 1$\,EeV and the anisotropy measured at Earth is expected to significantly differ from the averaged values presented in the previous section.

\section{Conclusions and Perspectives}
\label{Conclusions}

In this work we studied the consistency of a transition from Galactic to 
extragalactic CRs with existing anisotropy limits as a function of energy 
above $E=10^{18}$\,eV. The diffusion approximation predicts a dipole anisotropy 
$\delta=-3D_{ij}\nabla_j\ln(n)$ increasing with energy, since both the 
diffusion tensor $D_{ij}$ and the relative CR gradient $\nabla_j\ln(n)$ 
increase with energy. However, this approximation becomes unreliable
at ${\cal O}(E/Z)\sim 10^{16}$\,eV, and therefore we studied the 
propagation of  CRs in the Galactic magnetic field directly by backtracking trajectories.
We simulated the turbulent magnetic field on nested grids which allows one to
include turbulent field modes $\Bk$ with arbitrary small wave-lengths.
For the regular Galactic magnetic field we used up-to-date models from 
Ref.~\cite{Pshirkov:2011um}. Because the global structure of the GMF is still
rather uncertain, we studied the dependence of the resulting anisotropy on the 
magnetic field parameters such as its strength $B_0$, scale height $z_0$,
correlation length $L_{\rm c}$ and exponent $\alpha$ of its power-spectrum.
We also examined the dependence of our results on the width and height
of the disk in which sources are located.

The main results of this study are presented in the Figs. 3--5. They show
that the anisotropy mostly depends on the amplitude $B_0$ of the magnetic field in the disk. 
As our main conclusion from this study, we found that existing anisotropy limits 
are not compatible with light (proton) and intermediate (CNO) nuclei of Galactic origin as dominant contribution
to the CR flux above 1\,EeV. By contrast, Galactic iron nuclei as CR primaries are 
consistent with the existing limits even up to 10--20\,EeV, if the strength 
of the turbulent field is as large as $B_{\rm rms}\sim 8\,\mu$G.
This finding implies that determining the chemical composition of the CR flux
around $10^{18}$\,eV settles also the question of the transition energy
between Galactic and extragalactic component: As light nuclei at this
energy are not sufficiently isotropized, they have to be extragalactic.
Therefore the fast increasing proton contribution indicated by the 
KASCADE-Grande collaboration between $10^{17}$\,eV and $10^{18}$\,eV suggests
the beginning of an extragalactic component.

We also studied qualitatively the dependence of the anisotropy on the 
effective density of sources, see Figs.~6--8. The average escape time of 
iron nuclei with 10\,EeV energy from the Galaxy is $\sim 10^{5}$\,yr.
Assuming for magnetars a rate of $10^{-3}/$yr, the effective density 
of magnetars as sources of CR at 10\,EeV is $\sim 100$/Galaxy. Thus magnetars satisfy the anisotropy constraint and can be natural candidates for the sources of the high-energy end of the Galactic CR flux in the scenario where the transition from Galactic to extragalactic cosmic rays occurs at the ankle, provided they are
able to accelerate iron up to few$\;\times 10^{18}$\,eV.

In summary, we conclude that models with a transition from Galactic to 
extragalactic cosmic rays around the ankle are consistent with the existing 
anisotropy limits if the composition of Galactic cosmic rays at 
$E \gtrsim 10^{18}$\,eV is dominated by heavy nuclei. In contrast, if the chemical composition at these energies turns out to be light or intermediate, a transition at the ankle would be very strongly disfavoured.

\acknowledgments

We thank Venya Berezinsky and Martin Pohl for useful comments and discussions.
GG acknowledges support both from the Research Council of Norway through an Yggdrasil grant, and from APC Paris laboratory (France). The works of GG and GS are supported by the Deutsche Forschungsgemeinschaft through the collaborative research centre SFB 676. GS acknowledges support from the State of Hamburg through the Collaborative Research program ``Connecting Particles with the Cosmos'', from the ``Helmholtz Alliance for Astroparticle Phyics HAP'' funded by the Initiative and Networking Fund of the Helmholtz Association.

\appendix
\section{Validitation of the nested grid method}

In this appendix, we verify the validity of the new method we propose to generate turbulent magnetic fields on nested grids, see Section~\ref{Method_TF}. We have reproduced the earlier results of Refs.~\cite{Casse:2001be,DeMarco:2007eh}, and present below our computations for the diffusion coefficient versus the Casse \textit{et~al.} ones~\cite{Casse:2001be}.

\begin{figure}
\begin{center}
\includegraphics[width=0.49\textwidth]{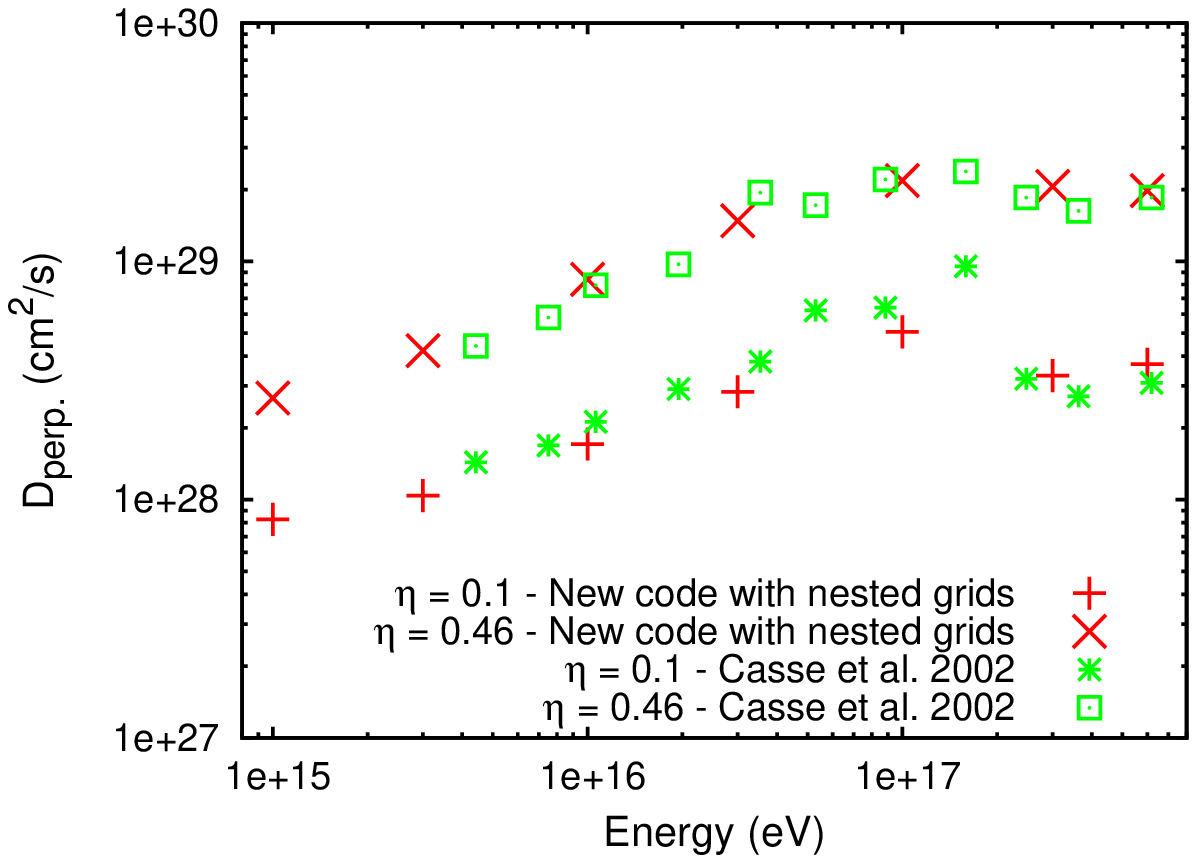}
\includegraphics[width=0.49\textwidth]{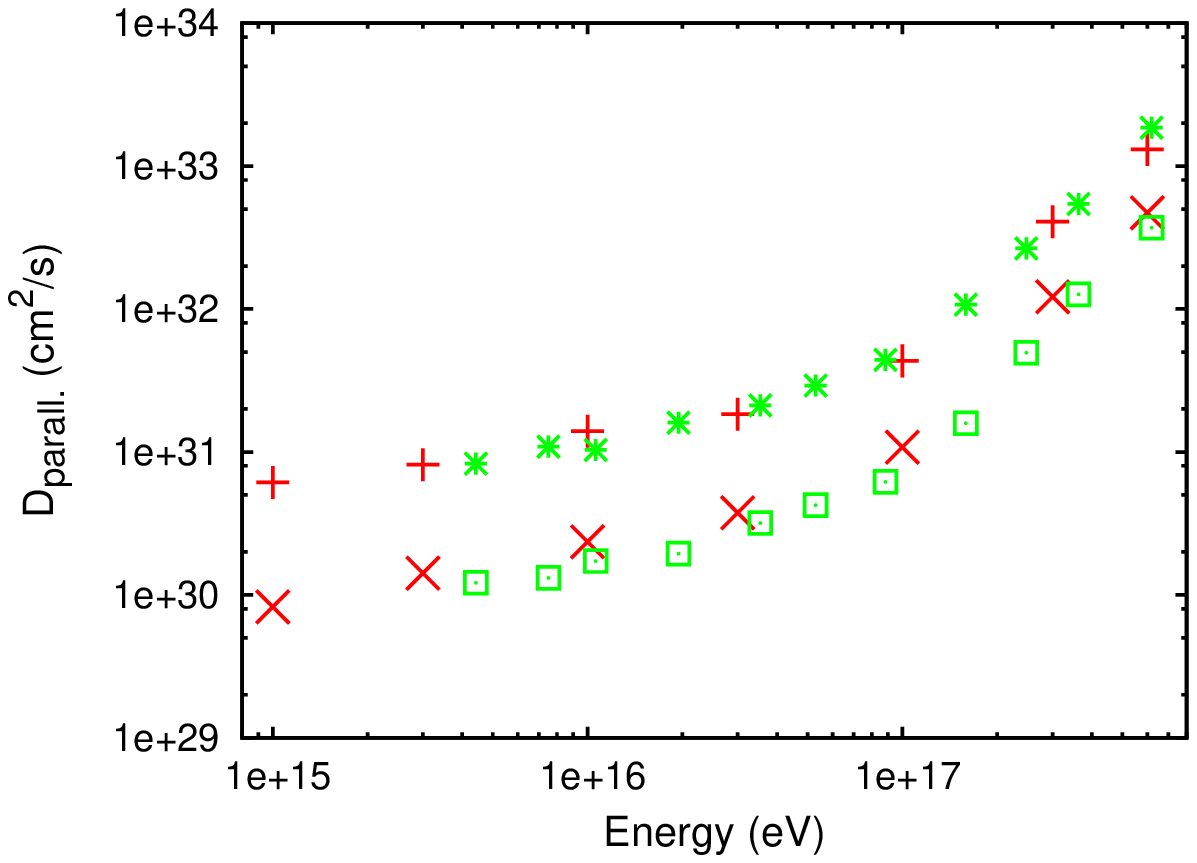}
\end{center}
\caption{Computations of the perpendicular \textit{(left panel)} and parallel \textit{(right panel)} diffusion coefficients for CR protons with energies $E=1-600$\,PeV, and for two different levels of turbulence $\eta = 0.1$ and $\eta = 0.46$, $\alpha = 5/3$ and $L_{\max}=150$\,pc for the turbulent magnetic field. $4\,\mu$G for the magnetic field strength. Red symbols for our results and green ones for the Casse \textit{et~al.} results~\cite{Casse:2001be}.}
\label{PerpParall}
\end{figure}

In Fig.~\ref{PerpParall}, we present the numerical results for the parallel (right panel) and perpendicular (left panel) diffusion coefficients for 1\,PeV to 600\,PeV CR protons diffusing in a field containing both a regular and a turbulent component. Let us denote $B_{\rm reg}$ the strength of the regular component, and $B_{\rm rms}$ the root mean square strength of the turbulent one. Two levels of turbulence are used $\eta = 0.1$ and $\eta = 0.46$, with $\eta = B_{\rm rms}^2 /(B_{\rm reg}^2 + B_{\rm rms}^2)$. The magnetic field strength is set to $4\,\mu$G. For the turbulent component, we take a Kolmogorov spectrum ($\alpha=5/3$) with $L_{\max}=150$\,pc and $L'_{\min}=0.1$\,pc. Therefore, $L'_{\min}$ is smaller than the Larmor radius $r_{\rm L}$ for all energies. We average over a few turbulent field configurations and propagate 2000~protons.

Red symbols in Fig.~\ref{PerpParall} represent our results with the nested grid code introduced in Section~\ref{Method_TF}, and green symbols represent the results of Figs.~4 and~5 of Ref.~\cite{Casse:2001be} adapted to the values we use here for $L_{\max}$ and the magnetic field strength. The uncertainties of our values and those of Casse \textit{et~al.}~\cite{Casse:2001be} can be estimated from the fluctuations from one point to another compared to averaged behaviour of the diffusion coefficients. Both for $\eta = 0.1$ and for $\eta = 0.46$, our results reproduce very well those of Casse \textit{et~al.}. We did not report in Fig.~\ref{PerpParall} the results of Casse \textit{et~al.} below $E \simeq 4$\,PeV because they correspond to computations for CRs with Larmor radius smaller than the minimum size of turbulent magnetic field fluctuations in their grid. This stresses one of the advantages of our new method: There is no lower limitation on $L'_{\min}/L_{\max}$ because one can always add another grid with smaller spacing and smaller scales of the magnetic field fluctuations. Therefore one can safely explore low rigidities.

\begin{figure}
\begin{center}
\includegraphics[width=0.6\textwidth]{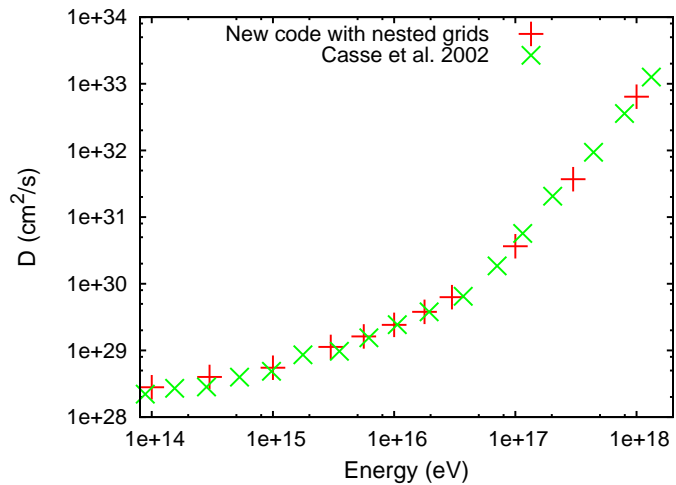}
\end{center}
\caption{Computations of the diffusion coefficient for CR protons with energies $E=100$\,TeV to 1\,EeV, in pure isotropic magnetic turbulence (no regular field); $\alpha = 5/3$, $B_{\rm rms}=4\,\mu G$ and $L_{\max}=150$\,pc for the turbulent field parameters. Red crosses for our results and green ones for the Casse \textit{et~al.} results~\cite{Casse:2001be}.}
\label{FullTurbulence}
\end{figure}

In Fig.~\ref{FullTurbulence}, we test our code for the case of 100\,TeV to 1\,EeV CR protons diffusing in a purely turbulent field, i.e.\ without any regular field, $B_{\rm reg}=0$. The parameters are the same as for Fig.~\ref{PerpParall}, except for $E=100-300$\,TeV where we take $L'_{\min}=0.01$\,pc to ensure that $L'_{\min}$ is smaller than $r_{\rm L}$. For computing time reasons, we keep $L'_{\min}=0.1$\,pc for $E\geq1$\,PeV. We use $B_{\rm rms}=4\,\mu$G and $\alpha=5/3$.

Red crosses in Fig.~\ref{FullTurbulence} correspond to our computations and green ones to those of Fig.~4 of Casse \textit{et~al.}~\cite{Casse:2001be} adapted to our values for $L_{\max}$ and $B_{\rm rms}$. One can see that also in this case the results are in very good agreement. As predicted theoretically, the diffusion coefficient is proportional to $E^{1/3}$ (respectively to $E^{2}$) at low (respectively high) rigidities. The computations of Casse \textit{et~al.} for pure turbulence were done with a field generated superposing Fourier modes. This enabled them to check such a wide rigidity range. However, such computations are significantly slower than ours, as discussed in Section~\ref{Method_TF}.

For the computations at $E\geq 1$\,PeV in Fig.~\ref{FullTurbulence}, we take two grids and the intermediate scale $L_2$ between the two grids is 5\,pc, which corresponds here to the Larmor radius of $E \approx 10$\,PeV protons. We have computed twice more values in the range $3-30$\,PeV than at other energies so as to check that this scale does not induce any artificial imprint in the results. As can be seen in the figure, the diffusion coefficient behaviour in this energy range is smooth and not affected by that intermediate scale.

\end{document}